\documentclass[graybox]{svmult}

\usepackage{mathptmx}       
\usepackage{helvet}         
\usepackage{courier}        
\usepackage{type1cm}        
\usepackage{makeidx}         
\usepackage{graphicx}        
\usepackage{multicol}        
\usepackage[bottom]{footmisc}

\usepackage{tikz}
\usepackage{todonotes}
\setuptodonotes{inline}
\usepackage{amssymb,latexsym,afterpage,calc,array,tabularx,amsmath,eepic,epic,subfigure, enumitem, hyperref}
\usepackage[capitalise]{cleveref}

\makeindex             

\usepackage{amssymb,latexsym,afterpage,calc,array,tabularx,amsmath,eepic,epic,subfigure}

\usepackage[T1]{fontenc}

\makeindex             

\begin{document}

\title*{Single-file pedestrian dynamics: a review of agent-following models}

\author{Jakob Cordes, Mohcine Chraibi, Antoine Tordeux, Andreas Schadschneider}
\authorrunning{J. Cordes, M. Chraibi, A. Tordeux,  A. Schadschneider} 

\institute{Jakob Cordes  \at  Institute of Advanced Simulation, Forschungszentrum J\"ulich GmbH, 52425 J\"ulich, Germany, \\ Institut f\"ur Theoretische Physik, Universit\"at zu K\"oln,
50937 K\"oln, Germany, \\ \email{j.cordes@fz-juelich.de}
\and Mohcine Chraibi \at Institute of Advanced Simulation, Forschungszentrum J\"ulich GmbH, 52425 J\"ulich, Germany, \\\email{m.chraibi@fz-juelich.de}
\and Antoine Tordeux \at School of Mechanical Engineering and Safety Engineering,
University of Wuppertal, 42285 Wuppertal, Germany \email{tordeux@uni-wuppertal.de}
\and Andreas Schadschneider \at Institut f\"ur Theoretische Physik and Institut f\"ur Physikdidaktik, Universit\"at zu K\"oln,
50937 K\"oln, Germany, \email{as@thp.uni-koeln.de}
}
%
%

\maketitle

\vspace{-2.0cm}
\begin{center}
\today
\end{center}

\vfill

\abstract*{Abstract}

Single-file dynamics has been studied intensively, both experimentally and theoretically. It shows interesting collective effects, such as stop-and-go waves, which are validation cornerstones for any agent-based modeling approach of traffic systems. Many models have been proposed, e.g. in the form of car-following models for vehicular traffic. These approaches can be adapted for pedestrian streams.
In this study, we delve deeper into these models, with particular attention on their interconnections. 
We do this by scrutinizing the influence of different parameters, including relaxation times, anticipation time, and reaction time.
 
Specifically, we analyze the inherent fundamental problems with force-based
models, a classical approach in pedestrian dynamics.
Furthermore, we categorize car-following models into stimulus-response and optimal velocity models, highlighting their historical and conceptual differences. 
These classes can further be subdivided considering the conceptual definitions of the models, e.g.\ first-order vs. second-order models, or stochastic vs. deterministic models with and without noise.
Our analysis shows how car-following models originally developed for vehicular traffic can provide new insights into pedestrian behavior. 
The focus on single-file motion, which is similar to single-lane vehicular traffic, allows for a detailed examination of the relevant interactions between pedestrians.

\section{Introduction}
\label{sec:Intro}

A plethora of models has been suggested for the description of pedestrian dynamics \cite{ChraibiKSS11,SchadschneiderCSTZ18,ChraibiTSS18}, often by introducing ad-hoc terms to existing models to putatively improve their applicability to certain situations. 
This has led to an ever-increasing zoo of models. 
Our objective is to systematize the plethora of pedestrian dynamics models by scrutinizing their interrelationships and comprehending the unique parameters they employ.
We will focus on continuous models and single-file pedestrian motion, as it is well studied and empirical data is available for comparison and model validation~\cite{BoltesZTSS18,Paetzke2022,ziemer2016,huang2018a,Portz2011,Jelic2012}. 
It turns out that this will also allow to gain insight into the role of some parameters that are specific for certain classes. 

The models discussed in the following are continuous in space and time. The dynamics are defined in terms of systems of first or second order differential equations. 
Some surprising connections between different models exist, which may be discovered through examination of limiting procedures for certain parameters. This will provide a more comprehensive understanding of the dynamics of the models as well as the significance of the parameters in the models.

After a brief review of the essential properties of single-file pedestrian motion, based on simulations as well as empirical studies, we will provide a short introduction to the popular class of force-based models. 
In the last years, it has been recognized that the force-based model class has inherent issues that are not related to the numerical methods used to solve the  equations. 
These issues stem from the strong inertia effects present in the model.
The inclusion of strong inertia effects in many pedestrian models is surprising since it is widely acknowledged that inertia plays a much smaller role in pedestrian motion compared to vehicular traffic. 
Despite this understanding, the role of inertia is not fully considered in the development of these models. 
In contrast, in models for highway traffic the situation is just the opposite. Here often first-order models are used where inertia effects are generally much less important. Inertia effects are then introduced into the models implicitly via different mechanisms. We try to shed some light on this apparent paradox by considering the relation between various models in more detail.

We will provide an overview of various models for vehicular traffic in single-file scenarios, such as one-lane roads, and their pedestrian counterparts.
This approach has a long history which will be summarized briefly.
We will then investigate the relationships between these models in more detail, allowing us to classify the approaches and identify the key elements of each class.
One of the central aspects is the relation between various timescales that have been introduced as model parameters, e.g.\ reaction time, relaxation time, anticipation time, etc. 
We try to elucidate the connections between these timescales and discuss how they impact the dynamics of the models from a broader perspective. 
This will help to uncover the various concomitant effects on the dynamics.

Stability analysis, both linear and nonlinear, is a key tool for investigating car-following models. We present a comprehensive overview of the historical background and provide a general understanding of the types of dynamics that can be observed in car-following models.
Finally, we will examine the impact of ``noise'' in these models, and demonstrate how it can lead to the formation of stop-and-go waves through a different mechanism than the typical instability mechanism.


\section{Single-File Motion}

In \emph{single-file motion}, agents move in a line without passing each other. 
This simplifies the theoretical description enormously because the motion is basically one-dimensional and the order of the agents remains unchanged. In the case of pedestrian dynamics, the interaction between the pedestrians is the dominating factor for the dynamics since other factors like route-choice are not relevant for single-file motion.
The strength of the interaction depends mostly on the distance between pedestrians and can often be restricted to nearest neighbours. 
Therefore this scenario is similar to car traffic on a single-lane road which allows to adapt ideas from vehicular traffic modeling. 
However, some care should be taken since vehicular and pedestrian traffic differ in certain aspects.

Another important reason to focus on single-file motion is the fact that good quality empirical data exist which allows a detailed comparison with model predictions. Furthermore, the simplicity of the single-file scenario is a good starting point for a better understanding of the intrinsic problems encountered in second order models. 


\subsection{Stop-and-Go Waves}
\label{sub_stopgo}

The terminology \emph{stop-and-go waves} has been introduced in the end of the 1970s \cite{duckstein1967control} for vehicles in a tunnel. Related terminologies are \emph{phantom jams} \cite{sugiyama2008traffic}, which is often used in popular context, or \emph{soliton} or \emph{kink} \cite{komatsu1995kink,nagatani1998modified} in mathematical literature. Another related concept is \emph{phase separation}, a terminology often used by physicists. Phase separation refers to the fact that two (or more) qualitatively different states coexist in different spatial regions at the same time. 
In traffic, this typically results in a region with moving vehicles and a region with stopped vehicles, known as a traffic jam.

\begin{figure}
\begin{center}
    \includegraphics[width= 0.5\textwidth]{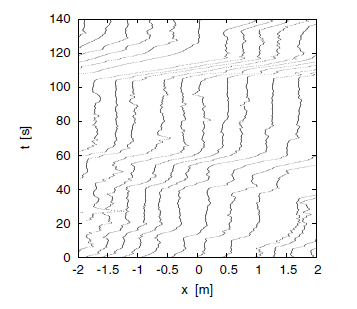}
    \includegraphics[width= 0.43\textwidth]{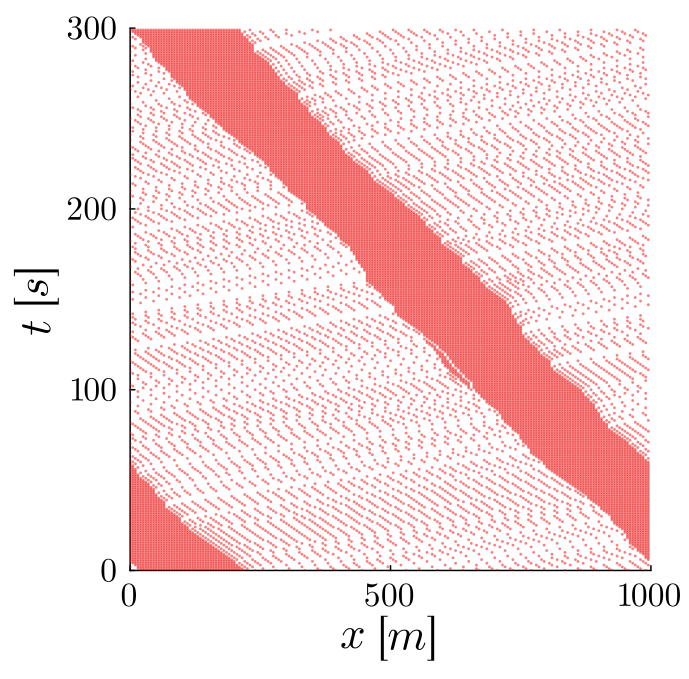}
    \vspace{-2mm}
    \caption{Left: Stop-and-go waves in trajectories from a single-file experiment with $N=70$ pedestrians
    (from \cite{seyfried2010phase});
    Right: Trajectories showing phase separation into a jam and a free-flow phase (from simulations of the VDR model \cite{BarlovicSSS98}).}
    \label{fig-stopngo}
\end{center}
\end{figure}

Stop-and-go waves in traffic flow have been studied for several decades \cite{herman_TrafficDynamicsAnalysis_1959,KernerR97,OroszWilsonSS} (see \cite{Chowdhury2000,Kerner2004,schadschneider_StochasticTransportComplex_2010} for reviews). Instead of forming homogeneous states, congested flows often self-organise into waves of slow and fast traffic also known as \emph{stop-and-go traffic} (\cref{fig-stopngo}, left). In cases where the formation of these waves can not be attributed to some external disturbance, also the terminology \emph{phantom jam} is used. Such cases have been observed in road, bicycle and pedestrian traffic \cite{zhang_UniversalFlowdensityRelation_2014}, both empirically and in experiments \cite{KernerR97,sugiyama2008traffic}. 

Stop-and-go waves move upstream, i.e. opposite to the motion of the agents, with a characteristic velocity. The formation of stop-and-go waves is an important collective phenomenon and benchmark test for models. It provides important information about the relevant interactions between the agents. Furthermore, it is also of practical relevance since it can impact safety, economy and comfort of transportation networks.

Due to its importance, stop-and-go behaviour has been investigated using different modeling approaches, i.e. with microscopic, mesoscopic (kinetic) and macroscopic models based on nonlinear differential systems (see for instance \cite{bando_DynamicalModelTraffic_1995,helbing_GasKineticBasedTrafficModel_1998,Colombo2003}), but also with discrete models like cellular automata. In models based on systems of differential equations the homogeneous equilibrium solutions,
where all agents move with the same velocity and headway can become unstable for certain values of the parameters. The solutions in the unstable regime are non-homogeneous and correspond to periodic or quasi-periodic solutions. 
In the latter case, potentially stop-and-go waves can occur after fine-tuning of the parameters.


\subsection{Phase Separation}
\label{sub_phsep}

Phase separation is a common and well-understood feature in vehicular traffic (see e.g.\ \cite{Chowdhury2000,schadschneider_StochasticTransportComplex_2010} and references therein) where two  distinct phases coexist: a jammed phase with high density and a free-flow phase with low density where cars move at their desired velocity with almost no interactions. The jam moves upstream with a characteristic velocity $v_\text{jam}$. This can be reproduced in models by using a \emph{slow-to-start} rule which reduces the outflow from a jam compared to the maximal flow in the system. It leads to spatially separated regions of non-interacting cars and thus to the free-flow phase and jams. Phase separation is related to the existence of metastable states in the fundamental diagram: for intermediate densities the flow is not a unique function of the density. The free flow branch is not stable and can spontaneously break down into a congested state, e.g.\ due to fluctuations in the velocities or an external disturbance. 
This is called a capacity drop and associated with a hysteresis loop~\cite{HallAG86}.

Note that phase separation is not exactly the same as stop-and-go waves which e.g.\ do not require a slow-to-start rule\footnote{E.g. the Nagel-Schreckenberg model~\cite{nagel_CellularModelFreeway_1992} shows stop-and-go waves, but no phase separation. The latter is observed only after a slow-to-start rule is added, e.g.\ in the VDR model \cite{BarlovicSSS98}.}. 
In the phase separated state, however, stop-and-go waves can form in the outflow from the large compact jam. 
In simulations with periodic boundary conditions, only one compact jam exists in the stationary state. Other jams will dissolve due to fluctuations. With open boundary conditions, several compact jams can coexist for some time, but they can dissolve and reform~\cite{AppertS01,BarlovicHSS02}.

Surprisingly, the structure of phase separation observed in pedestrian dynamics is different from that in vehicular traffic~\cite{seyfried2010phase}. Up to now, metastable states similar to those in vehicular fundamental diagrams have not been clearly identified in experiments with pedestrians. However, the trajectories of single-file pedestrian movement feature two separate phases: a jammed high-density phase, similar to that in highway traffic, and a phase of medium to high density with slowly moving pedestrians~\cite{seyfried2010phase}. The distance between pedestrians in the moving phase is small, not allowing them to move with their desired velocity, in contrast to vehicular traffic. Thus, both phases consist of interacting particles. The mechanism leading to phase separation therefore differs from a slow-to-start rule~\cite{EilhardtS14,EilhardtS15}.


\section{Force-Based Models}
\label{sec:SFM}

Besides cellular automata models, force-based (or acceleration-based) models are the most popular class of modeling approaches for crowd motion (see e.g.\ the reviews~\cite{ChraibiKSS11,SchadschneiderCSTZ18,ChraibiTSS18} and references therein). 
The appeal of these models lies in their resemblance to classical mechanics.
Pedestrians are considered as individual particles that interact with each other through various forces, both physical and ``social''.
These so-called \emph{social forces} are not based on the fundamental forces of physics, but rather on the observed deviation from a straight path when a pedestrian is approaching others. 
This deviation is seen as an acceleration which is the result of a force acting on the agent. 
The social force reflects a person's personal space and desire to avoid getting too close to others. 
Its concept also includes dynamical collision avoidance behaviors and extends the static relationships of proxemics, based on initiate, personal, social space, public space~\cite{Lewin1936,Lewin1943,Lewin51A}, to pedestrians in motion. 

Even though in some works it was attempted to deduce the description of the social forces from the trajectories of the pedestrians in simplified scenarios~\cite{Johansson2007,KrbalekHB18,Moussaid2009a}, 
the nature of the social force remains abstract. 
This characteristic of force-based models provides flexibility in their design, which has led to the proposal of several analytical expressions for these forces in the past.
This characteristic of force-based models provides flexibility in their design, which has led to the proposal of several analytical expressions for these forces in the past.
In general, one can distinguish between two main categories: ``algebraically decaying'' and ``exponential-distance'' models. 
In algebraically decaying models, the repulsive force between agents decays as some power of their distance. 
One typical choice is a force that is inversely proportional to the distance~\cite{YuW2005,Chraibi2010b,Shiwakoti2011,Lohner2010,Guo2010}. 
In contrast, in exponential-distance models, the magnitude of the repulsive force 
decays exponentially with increasing distance between the agents \cite{Helbing1995,Johansson2007,Moussaid2011,WangWang21}. 
The behavior of models in the two categories shows some differences.
For more details we refer the reader to~\cite{Chraibi2015a}, where the jamming transition was investigated for both types of interactions.

\subsection{Exponential-Distance Models}
\label{sub_SFM}

The \emph{social-force model (SFM)}~\cite{Helbing1995} is considered the first model that describes the movement of pedestrians by means of repulsive forces that decay exponentially with distance.
It is conceptually predated by the model of Hirai and Tarui (see Sec.~\ref{sub_GFBM}) which uses algebraic forces.
The SFM was first developed in the 1990s, and since then it has been widely used to study pedestrian dynamics in various settings. 
The SFM allows for the simulation of pedestrian behavior and the prediction of crowd movement in different scenarios, such as evacuation from a building. 
Over the years, it has undergone various modifications and enhancements to improve its accuracy in simulating pedestrian dynamics and to address its difficulties in describing certain scenarios (see~\cref{sub_problems}). 
Researchers have added physical forces (e.g. friction), improved numerical efficiency, and introduced mechanisms such as the ``respect mechanism''~\cite{Parisi2009} to improve the performance especially in crowded environments.

In~\cite{Moussaid2009a}, an approach for calibration of the SFM by using experimental measurements to adjust the repulsive force was proposed.
The updated repulsive force takes into account both the distance and the angle between the direction of movement and the vector that connects the positions of the two pedestrians.
This approach of considering both distance and angle dependencies is commonly used in other models, starting from~\cite{Hirai1975}, in order to account for the anisotropic behavior of pedestrians. 
Recent results show that rescaling of the social force model yields in a first-order Optimal Velocity model if inertia is neglected \cite{maury_CrowdsEquationsIntroduction_2018}. 
This over-damped limit of SFM (see also \hyperlink{app_oscillator}{Appendix~D}) can formally performed by letting the social forces becoming corrections to the optimal velocity. The exponential Weidmann pedestrian OV model \cite{weidmann1993transporttechnik} is then recovered if we restrict interactions to the next neighbor.

However, it is important to note that these measurements were primarily done with only two pedestrians and then extrapolated to a crowd with multiple individuals, which assumes that the superposition of forces can be applied in this context. 
This assumption, however, will turn out to be problematic and experimentally not sustainable (see~\cref{sub_problems}).
Additionally, different specifications of the repulsive force, such as circular or elliptical equipotential lines, are often used in SFM, first in~\cite{Helbing1995} and then improved in~\cite{Johansson2007}. 
However, for a one-dimensional analysis, which is our focus here, both of these specifications are equivalent. 
From a numerical point of view, an advantage of this subclass of models is  that it has no singularity at contact of two pedestrians.
For a detailed review, we refer to~\cite{Chen2017}.


\subsection{Algebraic-Distance Models}
\label{sub_GFBM}

The foundation of force-based models was established by Hirai and Tarui~\cite{Hirai1975}, where several ``social forces'' were introduced that rely on algebraic, piecewise linear relationships with the distance between pedestrians.
Based on the same principle, several models were developed, for instance the centrifugal \cite{YuW2005} and generalized centrifugal force model~\cite{Chraibi2010b}, where the effects of the relative velocities among pedestrians were investigated.

Another important aspect of pedestrian dynamics, \emph{volume exclusion}, was studied in more detail in another model variant introduced in~\cite{Seyfried2006}. 
Generically, following~\cite{Helbing2000}, pedestrians are represented by disks with constant radius. 
Observations have shown that the space requirement of pedestrians is not constant, but depends on the walking speed since e.g.\ the step length increases with increasing speed. 
This was taken into account in a model introduced by Seyfried et al. \cite{Seyfried2006}. 
Different from the two previously mentioned models, effects of the relative velocity were not considered here. 
Furthermore, Guo et al.~\cite{Guo2010} investigated a slightly different model with focus on navigation in two-dimensional space. Similar models introducing new features have been proposed in~\cite{Lohner2010} and \cite{Shiwakoti2011} with a constant added to the denominator of the repulsive force to avoid a singular force at short distances. 

A systematic investigation~\cite{ChraibiKSS11} of this model class has revealed that algebraic-distance models tend to be numerically unstable when compared to exponential-distance models, especially for high density scenarios. This can lead to problems during simulations. For a review and an extensive comparison, see~\cite{ChraibiKSS11}.


\subsection{Conceptual Problems}
\label{sub_problems}

The force-based approach has some fundamental issues that have not been fully addressed in the pedestrian dynamics field. For instance, unrealistic motion, 
such as oscillations, violations of the exclusion principle\footnote{The \emph{exclusion principle} states that the space occupied by a particle is not available for other particles.} (e.g. overlapping of agents and ``tunneling'' through other agents which leads to a reordering of agents in single-file motion), and acceleration to velocities larger than the desired velocity were highlighted in~\cite{Chraibi14,Kretz15}.
Furthermore, and probably related to these observations, it has been found that for obtaining realistic motion some physical parameters of the models have unrealistic values~\cite{Lakoba2005}. 
It should be emphasized that the mentioned effects are not a consequence of numerical problems~\cite{KoesterTG13}, e.g. in the discretization of the system of differential equations describing the motion, but are fundamental conceptual issues.

Another problem is related to the fact that social forces generically do not obey Newton's 3rd law, i.e. actio $=$ reactio. This raises fundamental issues, e.g. about the distinction between force and mass which relies on the 3rd law in classical mechanics.
Finally, the \emph{superposition of forces}, which states that the action of different forces on a body is given by the vectorial sum of the contributions, is not valid in many situations encountered in pedestrian dynamics~\cite{Seyfried19} and often leads to numerical problems (unrealistically high forces) and total forces pointing in the wrong direction. 
In single-file motion, the superposition of the driving force with an asymmetric repulsion can lead to unrealistic backward motion.

Some issues observed in force-based models can be attributed to excessive inertia effects, which can cause the model to exhibit damped oscillations instead of the desired overdamped behavior~\cite{schadschneider2020noise,cordes_Trouble2ndOrder_2020,Sticco20} (see App.~\ref{app_oscillator}).
In practice, the mass of the pedestrian is often used as the mass parameter, resulting in large inertia. To address this, a smaller ``effective'' mass could be introduced in simulations of second-order models, making the mass a parameter that needs to be determined through calibration.

Neglecting inertia effects reduces the dynamic equation for the damped harmonic oscillator to a 1st order differential equation. This observation has motivated the study of 1st order, i.e. velocity-based, models for pedestrian motion.
This is also reasonable since pedestrians can stop and turn almost immediately which is only possible if inertia effects are small.
For vehicular traffic, inertia effects are much more relevant than for pedestrian dynamics. Nevertheless, several 1st models that neglect inertia have been developed. From the previous arguments it seems worthwhile to investigate in more detail if they are suitable, possibly after some modifications, for pedestrian dynamics as well.

\section{Models for Single-File Motion}
\label{sec:models}

As discussed earlier, single-file motion is a simple scenario which allows to determine fundamental relations, e.g. between the pedestrian speed and distances to the neighbors, or the formation of stop-and-go waves.
Many researchers have conducted  experiments for single-file movement \cite{seyfried2005fundamental,cao2016pedestrian,zeng2021pedestrian,Paetzke2022}, and numerous high-quality empirical data is available (see \cite{BoltesZTSS18} and references therein) and the database \cite{_DatabaseJulich_}.

\subsection{Historical Overview}
\label{sub:hist_rew}

The first analyses of single-file motions date back to the 1950s with the pioneering works by Reuschel for the dynamics of road vehicles \cite{reuschel1950fahrzeugbewegungen}. 
In this model, the speed is instantaneously proportional to the distance gap. 
The car-following model is linear, of first order and without delay. 
A few years later, Pipes used a linear second-order car-following model with no delay for which the acceleration is proportional to the relative speed with the neighbor in front \cite{pipes_OperationalAnalysisTraffic_1953}.
The end of the 1950s saw the development of delayed linear models and stability analysis \cite{kometani1958stability,herman_TrafficDynamicsAnalysis_1959,chandler1958traffic,kishi1960traffic}, see \cref{sec:stab}. 
Non-linear car-following models emerged a few years later, at the beginning of the 1960s, with the first-order delayed exponential model by Newell \cite{newell1961nonlinear} and the general class of \textit{Stimulus-Response} follow-the-leader models by Gazis et al.\ \cite{gazis1961nonlinear}.
Although the pioneering model by Newell and the stimulus-response model are both non-linear, they are conceptually different. 
The model by Newell includes an equilibrium speed function depending on the spacing. 
Such a concept, initially introduced by Chandler to model constant time gap policy (\textit{California Code}) \cite{chandler1958traffic}, is central in modern car-following models. 
Indeed, most car-following models from the end of the 1990s rely on an explicit equilibrium speed-spacing function, generally related as the optimal velocity function. 
Prominent examples are the \textit{Optimal Velocity} model \cite{bando_DynamicalModelTraffic_1995}, the \textit{Intelligent Driver} model \cite{treiber_CongestedTrafficStates_2000}, or the \textit{Full Velocity Difference} model \cite{jiang_FullVelocityDifference_2001} 
(see \cref{sub:Cate} for details).

\begin{figure}[!ht]
\begin{center}\vspace{-2.5mm}
    \input{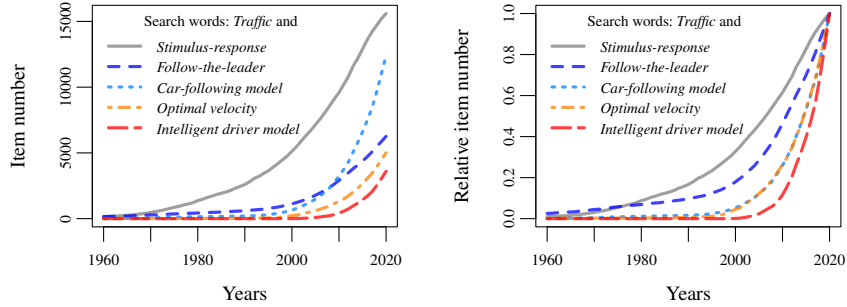}\vspace{-5.5mm}
    \caption{Numbers of items (left panel) and relative numbers of items (right panel) on Google Scholar for specific search words related to single-file motion models. Requests done on 22.08.2022.}
    \label{fig_RefCount}
\end{center}
\end{figure}

The numbers and relative numbers of items on Google Scholar for the search words \textit{Traffic} and \textit{Stimulus-response}, \textit{Follow-the-leader}, \textit{Car-following model}, \textit{Optimal velocity}, or \textit{Intelligent driver model} are presented in \cref{fig_RefCount}.
They allow for appreciating the terminology tendencies and the introduction of single-file motion concepts in traffic engineering during the last sixty years. 
The concept of \textit{stimulus-response} prevails in item number since the terminology is borrowed from psychology, where it has long been established as fundamental. 
The original terminology \textit{follow-the-leader} seems to be gradually declining in favor of \textit{car-following}. 
The \textit{optimal velocity} family of models becomes popular in the early 2000s.
Note that the terminology \textit{Single-file motion} or \textit{Single-file movement} is relatively recent and, until now, poorly established. 
As of 23.01.2023, it counts only 980 items on Google Scholar (156 in the year 2000 and 381 in 2010), including 340 with the additional search words \textit{Pedestrian} or \textit{Pedestrians}, and 592 with the words \textit{Cell} or \textit{Cells}. 
Interestingly, the terminologies \textit{Single-file motion} or \textit{Single-file movement} and \textit{Stimulus-response}, \textit{Follow-the-leader}, or \textit{Car-following} have only 68 items as of 23.01.2023! 
The analysis of pedestrian single-file motion is relatively young and, up to now, not connected to the rich following theory of traffic engineering.

Certain car-following models are used as pedestrian models for single-file motion, e.g., the stimulus-response model in \cite{Lemercier2012} or the optimal velocity model in \cite{Moussaid2011,Kuang2012}.
Other pedestrian single-file models are derived from two-dimensional force-based models \cite{Portz2011,Chraibi2015a}. 
It is worth mentioning that, in one dimension, the social force model corresponds to a multi-anticipative optimal velocity model.  
Indeed, the exponential pedestrian optimal velocity model by Weidmann \cite{weidmann1993transporttechnik} is recovered in one dimension by considering the social forces as corrections to the optimal velocity function. 
The vision field of the social force model then makes the interaction totally asymmetric. 
The resulting following model is a special form of the multi-anticipative car-following model by Lenz et al.\  \cite{lenz_MultianticipativeCarfollowingModel_1999} with short-range interaction (see, e.g., \cite{kretz2020analytical} for technical details).
The problems of agent overlapping, oscillation, and lack of body exclusion in the social force model pointed out in \cref{sub_problems} (see also  \cite{ChraibiKSS11,Chraibi14,Kretz15,cordes_Trouble2ndOrder_2020,Sticco20}) can be related in one dimension to drawbacks of the optimal velocity car-following  model. 
Indeed, the OV model requires fine-tuning of the parameters to avoid collisions, backward movements, and oscillations \cite{tordeux2014collision}.
Such unrealistic single-file behavior and shortcomings can be corrected using extended models including relative speed terms, such as models based on the time-to-collision, or using first-order optimal velocity models. 
These model classes will be discussed in more detail in the classification presented in \cref{sub:Cate}.

\subsection{Categorizing Following Models}
\label{sub:Cate}

As discussed in Sec.~\ref{sub:hist_rew}, many different types of following models exist. In the following we try to subdivide these models into different classes which then will help us to understand the relations between the models better. Without loss of generality we consider here only interactions with the nearest neighbor in front.
The general case of interactions with more than one predecessor will be discussed in~\hyperlink{AppA}{Appendix A}, see \cref{ModAccK}.

\begin{figure}[!ht]
\begin{center}
    \begin{tikzpicture}[thick]

\tikzset{
	particle/.style={
		circle,
		fill=black!25,
		draw = black!25,
		minimum size=#1
		},
	particle/.default = 25pt
	}
	
	\coordinate (z) at (-1.0,0);
	
	\coordinate (a) at (-0.0,0);
	\coordinate (b) at (3,0);
	\coordinate (c) at (6,0);
	\coordinate (d) at (8,0);

	\coordinate (x) at (9,0);
	
	\draw[dotted]
	 	(z) -- (a)
	 	(a) -- (b) 
	 	(c) -- (d) 
	 	(d) -- (x);	 	

	\path
	 	(a) node[particle]{} node[below =15pt] { $i-1$} 
		(b) node[particle]{} node[below =15pt] { $i$}
		(c) node[particle]{} node[below =15pt] { $i+1$}
		(d) node[particle]{} node[below =15pt] { $i+2$};

	\draw[-, draw=blue!]
	 	(b) -- node[above] { $\Delta x _{i}$} (c)
		(b) -- ++ (0, 0.1) -- ++ (0, -0.2)
		(c) -- ++ (0, 0.1) -- ++ (0, -0.2);
		
	\draw[->]
		(a) -- node[above] {$v_{i-1}$} ++ (1.5, 0.0);
		
	\draw[dashed]
	    (a) -- node[left, pos=0.8] {$x_{i-1}$} ++ (0.0, 0.8);

	\fill[]
		(a) circle [radius = 0.05];

	\draw[-]
		(d) ++ (-0.45, 0.6) -- node[above] {$\ell$} ++ (0.9, 0.0)
		(d) ++ (-0.45, 0.5) -- ++ (0.0, 0.2)
		(d) ++ (0.45, 0.5) -- ++ (0.0, 0.2);

	\path[->] (d) ++ (1.5, -0.5)  edge node[above] {$x$}  ++ (1.0, 0.0);

\end{tikzpicture}\vspace{-2mm}
    \caption{Illustration of an one-dimensional chain of self-driven particles. }
    \label{fig1}
\end{center}
\end{figure}
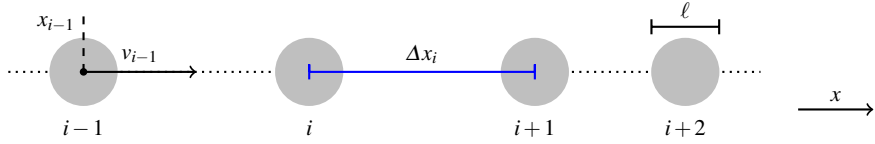

Since we focus on single-file motion, we assume a one-dimensional space with $N$ agents of size $\ell$, we denote their positions, velocities and accelerations by $x_i$, $v_i = \dot{x}_i$ and $a_i= \ddot{x}_i$, respectively (\cref{fig1}).
The agents are sorted in ascending order in the direction of motion, i.e. agent $i+1$ is the predecessor of agent $i$. in single-file motion this order will be preserved as overtaking is not possible. The distance between two agents $i$ and $j$ is $\Delta x_{ij} = x_j - x_i$, i.e. defined as the distance between the centers. For simplicity, $\Delta x_{ii+1}$ is denoted as $\Delta x_{i}$. Time-dependencies are omitted if all values are evaluated at the same time. 

Generically, all \emph{following models} then have the form
\begin{equation}
  \ddot x_i = A(\Delta x_i,\dot x_i,\dot x_{i+1})
  \label{ModAcc}
\end{equation}
for 2nd order models \cite{orosz2010traffic} or
\begin{equation}
  \dot x_i = V(\Delta x_i,\dot x_{i+1})
  \label{ModSpe}
\end{equation}
for 1st order models. The functions $A$ and $V$ are model-dependent.
For all models, there exists at least one equilibrium solution where the agents are evenly distributed in the system (i.e. $\Delta x_{i} = s$)  and move with the same speed $v_i =v$.  
Equilibrium speed $v$ and equilibrium spacing $s$ are determined by the conditions 
\begin{equation}
    A(s, v, v) = 0
  \label{EquModAcc}
\end{equation}
and
\begin{equation}
    V(s, v) = v\,,
  \label{EquModSpe}
\end{equation}
respectively. While for \emph{Stimulus-Response Models} this equilibrium solution is not unique, \emph{Optimal Velocity} models explicitly include a function such that the equilibrium solution is $(F(s), s)$. In some cases, to which we refer as \emph{Implicit Optimal Velocity models}, this function exists implicitly, but may be non-analytical. This categorization is sketched in~\cref{fig:Categories}.

\begin{figure}
\minipage{0.3\textwidth}
    \centering
    \tikzset{
	io/.style={
		fill=white
		},
	io2/.style={
		fill=white,
		text width = 4.15cm
		},
	ip/.style={
		rectangle,
		fill=blue!5,
		draw = blue!30
		},			
	op/.style={
		rectangle,
		fill=blue!5,
		inner sep = 8pt,
		minimum width = 3.5cm, 
    		minimum height = 1.2cm,
		draw = blue!80
		},
	cn/.style={
		rectangle,
		fill=blue!25,
		inner sep = 8pt,
		draw = blue!80,
		minimum width = 3.5cm, 
    		minimum height = 1.2cm
		},
	node distance = 10 mm
	}

\begin{tikzpicture}[ultra thick]

		\node[cn, align=center] (PC) at (-4.5, 1.5) {\textbf{Stimulus Response}};
		\node[op, align=center] (ANT1) at (-4.5, 0) {No unique eq. solution};
		\node[op, align=center] (ANTD) at (-4.5, -1.5) {Pipes \cite{pipes_OperationalAnalysisTraffic_1953}, \\ Chandler \cite{chandler1958traffic}, \\ Gazis \cite{gazis1961nonlinear}};
		
		\node[cn, align=center] (FD) at (-0.5, 1.5) {\textbf{Optimal Velocity}};
	 	\node[op, align=center] (OV1) at (-0.5, 0) {Explicit Function $F$: \\ $(F(s), v)$ is eq. solution};
 		\node[op, align=center] (OVD) at (-0.5, -1.5)  {Optimal Velocity \\ Model \cite{bando_DynamicalModelTraffic_1995},\\ FVDM \cite{jiang_FullVelocityDifference_2001}};

		\node[cn, align=center] (VO) at (3.5, 1.5) {\textbf{Implicit} \\ \textbf{Optimal Velocity}};
	 	\node[op, align=center] (OV1) at (3.5, 0) {$F$ exists implicitely, \\ e.g. non-analytical};
 		\node[op, align=center] (FVDMD) at (3.5, -1.5)  {Gipps \cite{gipps_BehaviouralCarfollowingModel_1981}, \\ Intelligent Driver \\ Model \cite{treiber_CongestedTrafficStates_2000}};

\end{tikzpicture}
\endminipage\hfill
\caption{The proposed categorization of car-following models (top row) with popular examples of each category (bottom row) and typical mathematical properties of the underlying dynamical equations (middle row).}
\label{fig:Categories}
\end{figure}

\begin{figure}
\minipage{0.5\textwidth}
    \centering
    \begin{tikzpicture}[ultra thick]

	\draw[->] (0,0) -- (4, 0) node[right]{$\rho$};
	\draw[->] (0,0.0) -- (0, 2.5) node[left]{$J(\rho)$};	
	
	\draw[-] (-0.05, 2) -- (0.05, 2) node[left]{\small $v_{max} \cdot \rho_c$};
	\draw[dotted, thin] (0.0, 2) -- (0.666, 2);
	\draw[dotted, thin] (0.666, 0.0) -- (0.666, 2);

	\draw[-] (0.666, -0.05) -- (0.666, +0.05) node[below]{\small $\rho _c$};
	\draw[-] (3.33, -0.05) -- (3.33, +0.05) node[below]{\small $1/\ell$};

	\draw[red] plot[domain=0.666:3.33] (\x, 2.5 * 1 - 2.5 * \x * 0.3 );
	\draw[red] plot[domain=0.0:0.668] (\x, 2.5 * \x * 1.2 );

\end{tikzpicture}
\endminipage\hfill
\minipage{0.5\textwidth}
    \centering
    \begin{tikzpicture}[ultra thick]

	\draw[->] (0,0) -- (4, 0) node[right]{$\rho$};
	\draw[->] (0,0.0) -- (0, 2.5) node[left]{$v(\rho)$};	
	
	\draw[-] (-0.05, 1.7*1.2) -- (0.05, 1.7*1.2) node[left]{\small $v_{max}$};
	\draw[dotted, thin] (0.666, 0.0) -- (0.666, 2);

	\draw[-] (0.666, -0.05) -- (0.666, +0.05) node[below]{\small $\rho _c$};
	\draw[-] (3.33, -0.05) -- (3.33, +0.05) node[below]{\small $1/\ell$};

	\draw[blue] plot[domain=0.666:3.33] (\x, 1.7 * 1 / \x - 1.7 * 0.3 );
	\draw[blue] plot[domain=0.0:0.668] (\x, 1.7 * 1.2 );

\end{tikzpicture}
\endminipage\hfill
\caption{A typical fundamental diagram for Optimal Velocity models.}
\label{fig:fund_dia}
\end{figure}
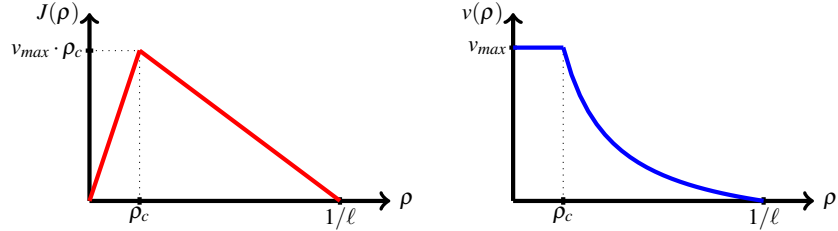

In \emph{Optimal Velocity models}, the existence of the described optimal velocity function $F(s)$ directly gives rise to the fundamental diagram (i.e. flow-density or speed-density relation) of the equilibrium solution. For a linear optimal velocity function with a maximal speed $v_{max}$ the typically observed shape is recovered, cf.~\cref{fig:fund_dia}. In the presence of instabilities, however, the fundamental diagram can not be directly inferred from the optimal velocity function. In practice, however, the difference is often very small.

\section{Generalized OV-Framework}
\label{sec:gen_ov}

The classification of Optimal Velocity models in~\cref{sub:Cate} is based on mathematical properties. It does not involve detailed behavioural assumptions about the agents and can include a broad variety of models. New models are often defined by adding \emph{ad-hoc} terms to existing models. Frequently these additional terms only have a small quantitative effect without changing the overall behavior. The following considerations might lead to a deeper understanding of this observation.

In this section, we try to show how and under which assumptions a class of Optimal Velocity models can be derived from a behavioural first principle, namely that agents keep a certain time-gap to their neighbors \cite{zhang_UniversalFlowdensityRelation_2014}.
By making the agents more human, i.e. adding \emph{anticipation} and a finite \emph{reaction} time, several well-known Optimal Velocity models are introduced and related to each other.
We will see that different timescales will become relevant here. These timescales have been introduced previously to capture certain aspects of pedestrian motion, in particular the desired time-gap, reaction or relaxation time, and anticipation time. 

In force-based models the \emph{relaxation time} $\tau _R$, or, more precisely, the speed-relaxation time, is usually implemented by a driving force $(v_\text{des}-v_i)/\tau _R$ where $v_\text{des}$ is the desired velocity. Neglecting other forces, this leads to an exponential relaxation of the walking speed to the desired velocity. As pointed out in \cite{johansson_ManyRolesRelaxation_2014}, in a $2$d scenario the same relaxation time also controls the dynamics of evasive motion when encountering obstacles or other pedestrians. The fact that the timescales for these different types of motion are identical can lead to problems. 
In OV models, the inverse relaxation time $1/\tau_R$ is interpreted as a sensitivity to deviations of the actual velocity from the optimal velocity \cite{bando_DynamicalModelTraffic_1995}. The difference between reaction, update and adaptation time has been analyzed in \cite{kesting_HowReactionTime_2008} for vehicular traffic. 
For pedestrian motion the time-gap is used in the \emph{Collision Free Speed model} \cite{tordeux_CollisionFreeSpeedModel_2016}, where the desired time-gap can be used to model the motivation in a crowd \cite{rzezonka_AttemptDistinguishPhysical_2022}. Introducing an anticipation time $\tau _A$ improves the realism of lane-formation in the model \cite{xu_AnticipationVelocitybasedModel_2021}.


\subsection{The Time-Gap and Optimal Velocity Models}

The typical \emph{time-gap} (in one-dimension) between agents $i$ and $j$  is
\begin{equation}
 \tau _{ij} = 
\begin{cases} 
 \dfrac{\Delta x_{ij} - \ell}{v_i} & \text{ for } v_i > 0 \\
 \infty & \text{ else,}
 \end{cases}
 \label{eq:time-gap}
\end{equation}
where it was assumed that $i$ is behind $j$, i.e. $x_i < x_j$. It denotes the time until agent $i$ would collide with agent $j$, if $j$ suddenly stops and $i$ keeps moving at a constant speed. Therefore, it can also be viewed as a \emph{time-to-collision}. 

In order to define a minimal model, we assume a behavioral first-principle. In particular, the agents want to move as fast as possible while keeping a time-gap greater or equal than the \emph{desired time-gap} $T$ to all other agents. This can be written as:
\begin{enumerate}
\item[]\emph{Agents choose their velocity $v_i$ such that $\tau _{ij} \geq T$ for all $j \neq i$ while at the same time maximizing $v_i$.} 
\label{principle}
\end{enumerate}
Note that, the agents can always fulfill this constraint by stopping, i.e. choosing $v_i = 0$.
More formally, the model can be written as
\begin{equation}
v_i = \max_{j \neq i} \left\{ v \in  \mathbb{R}  \, \Big| \frac{\Delta x_{ij}-\ell}{v} \geq T\right\},
    \label{principle_forumla}
\end{equation}
where the domain of $v$ might be restricted by a minimal or maximal velocity. The model defined by \cref{principle_forumla} in principle includes interaction with all other agents. However, due to the limited human perception, the interaction between pedestrians is usually restricted to those in front. Furthermore, by definition, the ordering is preserved in single-file motion, i.e. $x_{i+1}(t) > x_i(t)$. Here we leave aside the frequent problem of tunneling, which leads to a destruction of this order and can give rise to numerical problems and artifacts (see~\hyperlink{AppC}{Appendix C} for more details).

With these assumptions, \cref{principle_forumla} simplifies to an \emph{Optimal Velocity} model which only includes an interaction with the nearest neighbor in front, i.e.
\begin{equation}
  \tag{OV, $1^{st}$}
    v_i = \frac{\Delta x_{i} -l} {T}.
\label{ov_model}
\end{equation}
This minimal model is the original first-order Optimal Velocity model introduced by Reuschel \cite{reuschel1950fahrzeugbewegungen} and Pipes \cite{pipes_OperationalAnalysisTraffic_1953} with a linear OV function. 


\subsection{Reaction and Anticipation Time}

An important constraint in human traffic is the \emph{reaction time}, by which human reactions to a change in the environment are delayed. 
A finite reaction time can be included by adding $\tau _R$ as a delay to the corresponding quantities, i.e. evaluating them at a later time.
The corresponding delayed first-order model of \cref{ov_model} reads
\begin{equation}
\tag{OV, $del$}
v_i(t+\tau_R) = \frac{\Delta x_{i}(t) -l} {T},
\label{delayed_first_order_OV}
\end{equation}
which is a delay differential equation. Second-order models can be obtained by employing a Taylor expansion for small $\tau _R$ as
\begin{equation}
v_i(t+\tau _R) \simeq v_i(t) + \tau _R \cdot a_i(t),
\label{reaction_apprx}
\end{equation}
where $a_i = \dot{v}_i$. Applying \cref{reaction_apprx} to \cref{delayed_first_order_OV}, results in
\begin{equation}
\tag{OV, $2^{nd}$}
a_i = \frac{1}{\tau _R} \left( \frac{\Delta x_{i} -l} {T} - v_i\right),
\label{2nd_order_OV}
\end{equation}
the second-order OV model \cite{bando_DynamicalModelTraffic_1995}. The Taylor expansion in \cref{reaction_apprx} considerably simplifies the model \cref{delayed_first_order_OV}. Note that this linearisation can have consequences on the stability properties of the models which will be discussed in \cref{sec:stab}.
In order to allow a higher and more comfortable flow, and to account for the reaction time, humans \emph{anticipate} (possible) changes of the current situation. The time-gap \cref{eq:time-gap} can be interpreted as anticipating a worst case scenario, i.e. that the preceding agent suddenly stops. Another typical anticipation strategy is to assume that the preceding agent keeps on moving with a constant velocity. At the level of a time distance this gives rise to the time-to-collision which will be discussed below. In our case, a forecast of future positions based on the relative velocity can be implemented as
\begin{equation}
\Delta x_{i}(t+\tau _A) \simeq \Delta x_{i}(t) + \tau _A \cdot \Delta v_{i}(t),
\label{ant_apprx}
\end{equation}
where $\tau _A$ is the \emph{anticipation time}. Adding $\Delta x_{i}(t+\tau _A)$ as a forecast, i.e. as true knowledge of the future positions, to \cref{delayed_first_order_OV} changes its delay simply to $\tau _R - \tau _A$. This is an example where the introduction of an additional parameter does not lead to changes in the fundamental behavior of a model.

Even though \cref{ant_apprx} and \cref{reaction_apprx} are very similar, their natural interpretation is different. In particular the former is an assumption about the behaviour of the agents, whereas the latter is an approximation to obtain a simpler model. 

If anticipation is added as in \cref{ant_apprx} to the model \cref{ov_model}, one obtains
\begin{equation}
\tag{ANT, $1^{st}$}
v_i =  \frac{\Delta x _{i} - l}{T+\tau _A} + v_{i+1} \cdot \frac{\tau _A}{T+\tau  _A},
\label{model_ant_first_order}
\end{equation}
which is a specific version of the time-to-collision model proposed in \cite{cordes_TimeToCollisionModelsSingleFile_2022}. 
A reaction time $\tau _R$ can be included as
\begin{equation}
\tag{ANT, $del$}
v_i(t+\tau _R) =  \frac{\Delta x _{i}(t) - l}{T+\tau _A} + v_{i+1}(t) \cdot \frac{\tau _A}{T+\tau  _A}.
\label{model_ant_first_order_delay}
\end{equation}
Employing \cref{reaction_apprx} yields 
\begin{equation}
\tag{ANT, $2^{nd}$}
a_i = \frac{1}{\tau _R} \left( \frac{\Delta x _{i} - l}{T+\tau _A} + v_{i+1} \cdot \frac{\tau _A}{T+\tau _A} - v_i \right).
\label{model_ant_2nd_order}
\end{equation}
Anticipation can also be included in \cref{delayed_first_order_OV}, i.e.
\begin{equation}
\tag{FVDM, $del$}
v_i(t+\tau_R) =  \frac{\Delta x _{i}(t) - l}{T} + \Delta v_{i}(t) \cdot \frac{\tau _A}{T}.
\label{FVDM_delay}
\end{equation}
Together with \cref{reaction_apprx} this leads to
\begin{equation}
\tag{FVDM, $2^{nd}$}
a_i = \frac{1}{\tau _R} \left( \frac{\Delta x _{i} - l}{T} + \Delta v_{i} \cdot \frac{\tau _A}{T} - v_i \right),
\label{FVDM}
\end{equation}
the well known \emph{Full-Velocity-Difference model} \cite{jiang_FullVelocityDifference_2001}. 
In the models \cref{model_ant_first_order_delay} and \cref{model_ant_2nd_order} anticipation is added before a finite reaction time, while in \cref{FVDM_delay} and \cref{FVDM} we first added a finite reaction time and then anticipation.
This boils down to the question if agents know their own velocity without a delay (as in the former) or with a delay (as in the latter).
This leads to considerable differences in the stability properties which lie beyond the scope of this review.


\subsection{Summary of Model Relations}

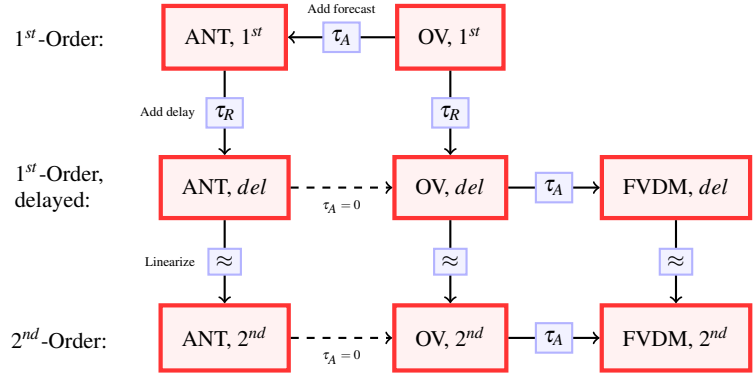
\begin{figure}
\minipage{0.6\textwidth}
    \centering
    \tikzset{
	io/.style={
		fill=white
		},
	io2/.style={
		fill=white,
		text width = 4.15cm
		},
	ip/.style={
		rectangle,
		fill=blue!5,
		draw = blue!30
		},			
	op/.style={
		rectangle,
		fill=red!5,
		inner sep = 8pt,
		draw = red!80
		},
	cn/.style={
		rectangle,
		fill=white,
		inner sep = 8pt,
		draw = black!
		},
	node distance = 10 mm
	}

\begin{tikzpicture}[ultra thick]
		

	 	\node[op] (OV1) at (-1.5, 0) {\ref{ov_model}};
 		\node[op] (OVD) at (-1.5, -2)  {\ref{delayed_first_order_OV}};
 		\node[op] (OV2) at (-1.5, -4) {\ref{2nd_order_OV}};

 		\node[op] (FVDMD) at (1.5, -2)  {\ref{FVDM_delay}};
 		\node[op] (FVDM) at (1.5, -4)  {\ref{FVDM}};
 		
		\node[op] (ANT1) at (-4.5, -0) {\ref{model_ant_first_order}};
		\node[op] (ANTD) at (-4.5, -2) {\ref{model_ant_first_order_delay}};
		\node[op] (ANT2) at (-4.5, -4) {\ref{model_ant_2nd_order}};
		
		\node[align=left] (O1) at (-6.7, -0) {$1^{st}$-Order:};
		\node[align=left] (OD) at (-6.7, -2) {$1^{st}$-Order, \\delayed:};
		\node[align=left] (O2) at (-6.7, -4) {$2^{nd}$-Order:};

 			
 		\path[->, bend left = 0, thick] (OV1) edge node[ip]{$\tau _R$} (OVD);
 		\path[->, bend right = 0, thick] (OV1) edge node[ip]{$\tau _A$} node[above = 6pt]{\tiny Add forecast} (ANT1);
 		
 		\path[->, bend right = 0, thick] (ANT1) edge node[ip]{$\tau _R$} node[left=8pt]{\tiny Add delay} (ANTD);
 		\path[->, bend right = 0, thick] (ANTD) edge node[ip]{$\approx$} node[left=8pt]{\tiny Linearize} (ANT2);
 		
 		\path[->, bend left = 0, thick] (OVD) edge node[ip]{$\tau _A$} (FVDMD);
 		\path[->, bend right = 0, thick] (OVD) edge node[ip]{$\approx$} (OV2);

		\path[->, bend left = 0, thick] (FVDMD) edge node[ip]{$\approx$} (FVDM);
		\path[->, bend right = 0, thick] (OV2) edge node[ip]{$\tau _A$} (FVDM);

		\path[->, bend left = 0, thick, dashed] (ANTD) edge node[below = 1.5pt, io]{ \tiny $\tau _A = 0$} (OVD);
		\path[->, bend right = 0, thick, dashed] (ANT2) edge node[below = 1.5pt, io]{ \tiny $\tau _A = 0$} (OV2);

\end{tikzpicture}
\endminipage\hfill
\caption{Chart of the generalized Optimal Velocity models.}
\label{fig:chart}
\end{figure}

The connections of the proposed models by the various time-scales is summarized in~\cref{fig:chart}. The derivation of Optimal Velocity models from the \emph{time-gap} leads to a loss of generality: In the typical formulation, the optimal velocity function $F(\cdot)$ is not specified but only constrained by few assumptions \cite{wilson_CarfollowingModelsFifty_2011}. 
From another perspective, however, it is more general: the Ansatz \cref{principle_forumla}, is neither restricted to a specific time-gap, a one-dimensional setting nor to circular shapes. If the time-gap in \cref{eq:time-gap} is formulated in two-dimensions, the collision-free speed model \cite{tordeux_CollisionFreeSpeedModel_2016} follows  directly from the same assumptions.

By changing the time-gap to the time-to-collision, single-file time-to-collision models can be defined~\cite{cordes_TimeToCollisionModelsSingleFile_2022}. 
In a two-dimensional setting this leads to \emph{Velocity Obstacle models} which were introduced for motion planning in complex environments \cite{fiorini_MotionPlanningDynamic_1998} and used to model pedestrian motion as well \cite{vandenberg_ReciprocalVelocityObstacles_2008, vandenberg_ReciprocalNBodyCollision_2011}. Note that these models are usually defined in discrete time, i.e. with an explicit \emph{update time} $\delta t$. In discrete time models this has to be considered as part of the model definition and a parameter which needs to be calibrated. In time-continuous models the update time comes from the (time) discretisation of the dynamical equations for their numerical treatment. It is therefore not a model parameter, but nevertheless very important: it should be small enough such that the numerical results are independent of it.

The time-scales introduced above, namely, the update time, the time-gap, the desired time-gap, the reaction time, and the anticipation time, have helped to uncover relations between various models. In order to further understand their meaning and effects, these models are investigated regarding their stability properties \cref{sec:stab} and the effect of noise \cref{sec:noise}.


\section{Stability Analysis}
\label{sec:stab}

Stability analysis is a technique used to determine the long-term behavior of perturbed systems, specifically their response to small deviations from the steady state. 
It allows to determine the conditions under which a system is stable or unstable. Stable systems return to their original state after a disturbance whereas the disturbance is amplified in unstable systems. 
The technique is applied to a mathematical model of the system, which is typically represented by a set of ordinary differential equations (ODEs) or delay differential equations (DDEs). 

Stability analysis is crucial in understanding the dynamics of single-file systems, such as stop-and-go waves and phase separation (see~\cref{sub_stopgo} and \cref{sub_phsep}). 
Many studies have shown that delays and large relaxation times in the dynamics tend to lead to instability and the formation of stop-and-go waves in nonlinear models. 
The stability analysis can be performed using different techniques, such as the Laplace transform, exponential Ansatz, and Hurwitz conditions for periodic and infinite systems by ODEs. 
Indeed, linear ordinary differential equation systems have polynomial  characteristic equations. 
Proving stability then amounts to show that these polynomials are Hurwitz, i.e.,  their roots are located in the left half-plane of the complex plane  \cite{frank1946zeros,kuo1966network}.
The stability analysis of delayed models is more complex because the system characteristic equation is no longer a polynomial, but an exponential-polynomial equation for which the Hurwitz conditions do not apply. 
The Hopf bifurcation method is particularly useful for determining critical stability borders of delayed systems and deducing the stable domain through continuity \cite{orosz2004global,gasser2004bifurcation,orosz2010traffic,tordeux2012linear}.

We present in the two next subsections a historical overview on the stability analysis of single-file motions.
the stability analysis of queueing movements.
As a complement, we provide in~\hyperlink{AppA}{Appendix A} general linear stability conditions for first- and second order ODE models.
General linear stability conditions for models defined by first- and second-order DDEs are given in \hyperlink{AppB}{Appendix B}.


\subsection{The prolific 1950s and early 1960s}

The stability of following models has been an issue for almost 70 years. First stability analyses of a single-file of agents date back to the beginning of the 1950s with the pioneering works of A.\ Reuschel and L.\ Pipes \cite{reuschel1950fahrzeugbewegungen,pipes_OperationalAnalysisTraffic_1953}. 
These seminal applications focus on the behavior of road vehicles using first \cite{reuschel1950fahrzeugbewegungen} and second \cite{pipes_OperationalAnalysisTraffic_1953} order models with no delay. 
A few years later, towards the end of the 1950s, delayed linear models were studied \cite{kometani1958stability,herman_TrafficDynamicsAnalysis_1959,chandler1958traffic,kishi1960traffic}, using, for the most part, the first and second order delayed models
\begin{equation}
\dot x_i(t+\tau)=\alpha(x_{i+1}(t)-x_i(t)-\ell)
\label{Kometani}
\end{equation}
and
\begin{equation}
    \ddot x_i(t+\tau)=\beta(\dot x_{i+1}(t)-\dot x_i(t)).
\label{Chandler}
\end{equation}

\begin{figure}
\bigskip
\minipage{0.33\textwidth}
    \centering \textbf{(a) Overdamped}
    \includegraphics[width=.95 \linewidth]{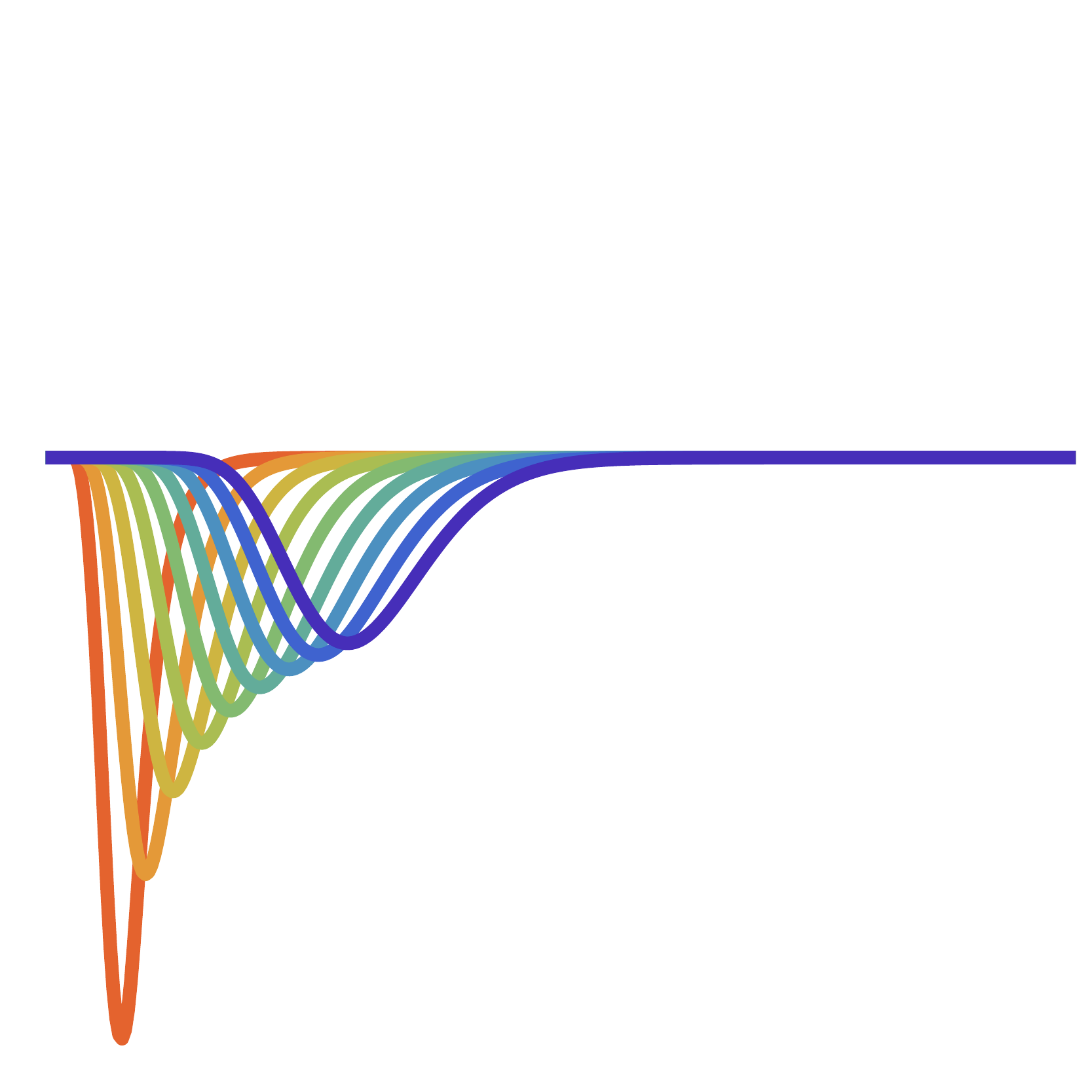}
\endminipage\hfill
\minipage{0.33\textwidth}
    \centering \textbf{(b) Oscillating}
    \includegraphics[width=.95 \linewidth]{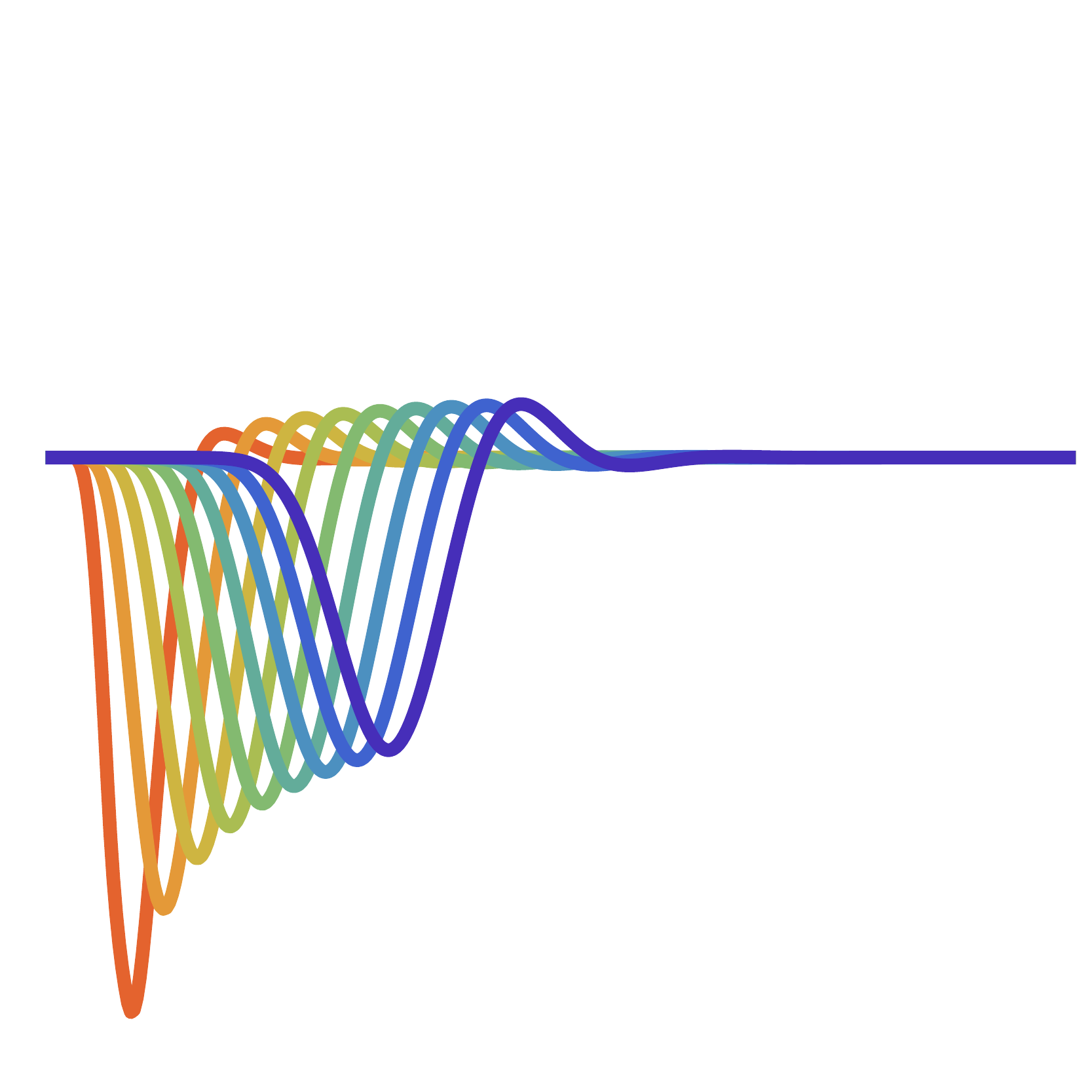}
\endminipage\hfill
\minipage{0.33\textwidth}
    \centering \textbf{(c) String Unstable}
    \includegraphics[width=.95 \linewidth]{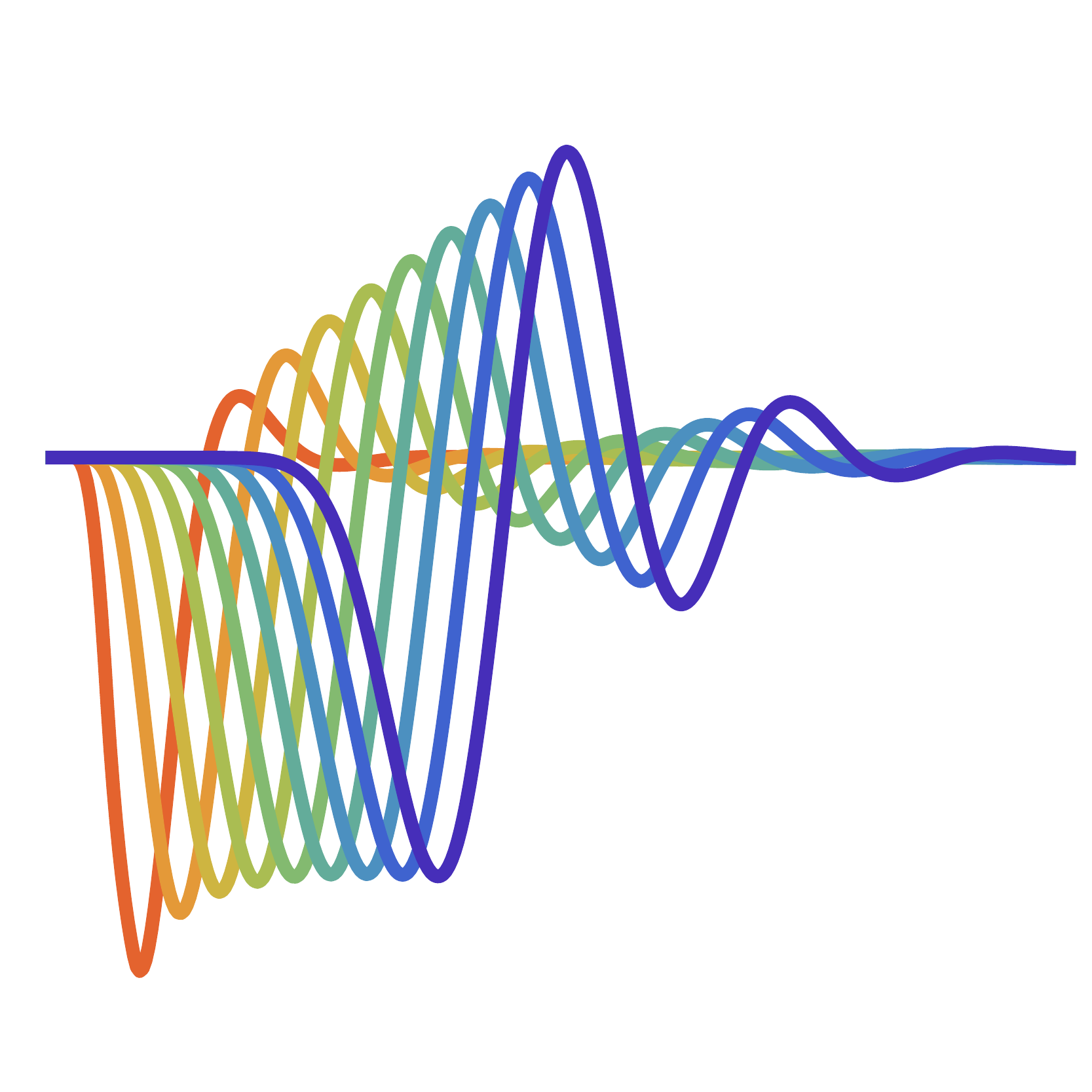}
\endminipage\hfill
\caption{Numerical solutions of a string of $10$ agents for \cref{FVDM}. All strings are platoon stable, i.e. oscillations attenuate for an individual agent.
While in (a) and (b) the perturbation is attenuating upstream, the agents 'overshoot' only in (b). In (c) The perturbation grows when moving upstream. The system is not string stable.}
\label{fig:stab}
\end{figure}

Most of the fundamental stability concepts were established during these very prolific years. 
From a methodological point of view, the first investigations were done using linear algebra \cite{reuschel1950fahrzeugbewegungen}. Then, Laplace transform to study a platoon of vehicles \cite{pipes_OperationalAnalysisTraffic_1953,kometani1958stability,herman_TrafficDynamicsAnalysis_1959} and Fourier analysis using an exponential Ansatz \cite{chandler1958traffic} quickly became the reference methods (see \cite{orosz2010traffic} for a survey). Conceptually, one distinguishes between
\emph{platoon stability}, for finite system with open boundaries, 
and \emph{string stability} for infinite systems and systems with periodic boundaries. 
The concepts of \emph{damped stability}, presenting oscillating behaviors, and \emph{overdamped stability}, without oscillations, were also established during this period \cite{kishi1960traffic} (see~\cref{fig:stab}).
An important physical insight was the negative impact of the delay on the stability of the flow. 
Unstable models amplify the perturbations resulting in the formation of traffic waves.
This was a considerable advantage over early fluid dynamics models \cite{lighthill1955kinematic,richards1956shock}, which are systematically stable.

\subsection{Resumption from the 1990s with nonlinear models}

Stability analysis then went through a period of decline, punctuated by the first analyses of multi-anticipatory models in the 1970s \cite{bexelius1968extended}, before rebounding in the late 1990s using nonlinear models.
Indeed, it has been early pointed out that the stability analysis of nonlinear models is more challenging \cite{gazis1961nonlinear}. 
In 1995, simulation results of the \textit{Optimal Velocity model} \cite{bando_DynamicalModelTraffic_1995}
\begin{equation}
\ddot x_i(t)=\frac1\tau _R\big(V(\Delta x_i(t))-\dot x_i(t)),
\label{OVM}
\end{equation}
with
\begin{equation}
V(x)=A\tanh(x-B)+C,
\end{equation}
showed that a nonlinear optimal velocity function $V(\cdot)$, a concept initially introduced at the beginning of the 1960s \cite{newell1961nonlinear,gazis1961nonlinear}, can have two stable stationary states according for certain parameter settings: a homogeneous state and a heterogeneous one with traffic waves propagating backward \cite{bando_DynamicalModelTraffic_1995}. 

These findings enabled to realise that nonlinear stability is the key feature explaining stop-and-go dynamics in highway traffic. 
Shortly afterwards, first investigations using the nonlinear reductive perturbation method allowed to derive the Korteweg/de Vries (KdV) \cite{komatsu1995kink} and later the modified KdV \cite{nagatani1998modified} macroscopic wave equation from the optimal velocity model. 
Certain nonlinear models can even have metastable states with multiple stationary dynamics depending on the initial conditions \cite{tomer2000presence}.
A few years later, several authors rigorously analysed nonlinear dynamics of different car-following models using Hopf bifurcation analysis \cite{orosz2004global,gasser2004bifurcation,orosz2010traffic,tordeux2012linear}. 
The results point out the existence of convective upstream, stationary, and convective downstream perturbation dynamics \cite{wilson_CarfollowingModelsFifty_2011}. 
These findings enabled to distinguish platoon stability, taking solely into account convective downstream perturbations, from string stability accounting for convective upstream, downstream, and stationary perturbation dynamics.
In parallel, stop-and-go waves have been observed experimentally with cars and pedestrians on a ring geometry \cite{seyfried2005fundamental,sugiyama2008traffic,seyfried2010phase}, or, more recently, with platoons of autonomous vehicles  \cite{Stern2018,gunter2020commercially,makridis2021openacc,CIUFFO2021}.


\section{The Effect of Noise}
\label{sec:noise}

Noise is a random disturbance that can affect the dynamics of a system. 
It represents the effect of factors that are not explicitly included in the model, but are still relevant. Usually, noise is modeled as white noise which is a time-independent random variable with a Gaussian distribution added to the model. 
Noise is not necessarily an unwanted perturbation, but can also have beneficial effects, such as preventing the system from getting stuck in a certain configuration (gridlock). However, additional computational effort is required to average over noise. 

Noise has to be distinguished from intrinsic stochasticity, e.g. in cellular automata models. Stochasticity is often an essential part of the dynamics and e.g. captures missing information about decision making. Therefore the deterministic limits of cellular automata models often show unrealistic behavior.

In this Section we will give more details on the role of various types of noise in models, starting with white noise before presenting colored noise models.


\subsection{White Noise Models}
\label{sub:noise}

Empirical data is always noisy, even for experiments under laboratory conditions. This noise is partly due to inaccuracies in the measurements, but can also be intrinsic to the system. However, this is usually difficult to distinguish.
In classical models, noise is usually added to the position and or velocity of the agents. 
Although usually not essential, it makes trajectories more realistic. It takes into account small fluctuations in the walking behavior that could be intrinsic or triggered by random external factors. As mentioned above, an important side effect is to prevent the system from freezing in a trivial configuration, e.g. a gridlock. This is often observed at unrealistically low densities in deterministic models (at least for certain initial conditions) and can already be avoided by relatively small noise. 

In continuous time, the white noise is introduced using independent Wiener processes $W_n$ increment. 
For instance, a first order model including an additive white noise reads
\begin{equation}
    d x_n(t)=V(\Delta x_n(t)) dt+\sigma dW_n(t),
\end{equation}
with $\sigma$ the noise volatility.
The noise being by definition centred, it does not affect the system linear stability \cite{treiber2017intelligent,friesen2021spontaneous}. 
More sophisticated models include state-dependent noises, see e.g. the stochastic traffic model by Yuang et al.\ for which the noise volatility $\sigma\propto\dot x_n$ is proportional to the speed \cite{yuan2019geometric,xu2020statistical}. 
Similar noise models rely Cox–Ingersoll–Ross processes \cite{ngoduy2019langevin}.
The noise volatility of this model class depends on the system state and its derivative may impact the deterministic linear stability condition. 

The effect of white noise in first order dynamics are relatively modest. 
Indeed, first order dynamics are over-damped and do not present oscillating behaviors \cite{schadschneider2020noise,cordes_Trouble2ndOrder_2020}. 
The role of the noise, even white, in second order and delay models including inertia can be much more preponderant and trigger, for instance, stop-and-go waves \cite{Tordeux2016b,treiber2017intelligent,friesen2021spontaneous}. 
Yet, stochastic second-order models can be recovered using colored noises in a first-order framework.


\subsection{Colored Noise Models}
\label{sub:correlatednoise}

White noises are uncorrelated in time and have a flat frequency power spectrum. 
Colored noise are time-dependent processes whose spectrum may be ether increasing (e.g., for blue or violet noise), or decreasing (for, e.g., pink of Brownian noise). 
The spectrum rise quantifies the noise fluctuation range and its inertia. 
Empirical time-series of the pedestrian speed show a power spectrum which is roughly proportional to $1/f^2$ for a large frequency range \cite{tordeux_NoiseInducedStopandGoDynamics_2019,Tordeux2016b}. 
This a typical characteristic of Brownian noise with large inertia.

The stochastic model introduced in \cite{Tordeux2016b} can be understood as a first order OV model with a correlated truncated Brownian noise $\epsilon_n(t)$ described by an Ornstein-Uhlenbeck process: 
\begin{equation}
\begin{split}
\dot{x}_n (t)= V(\Delta x_n(t))+\epsilon _n (t),\\\
d\epsilon _n (t) =-\frac{1}{\tau} \epsilon _n (t)dt + \sigma dW_n(t)\,.
\end{split}
\label{modeldef}
\end{equation}
$ V(d)=(d-l)/T $ is a linear OV function with desired time gap $T$ and $l$ the size of the agents. The $W_n(t)$ are Wiener processes. The (positive) parameters $\sigma$ and $\tau$ related to the noise can be interpreted as volatility and noise relaxation time, respectively.

This noisy first-order model is equivalent to \cref{FVDM} with a white noise and $\tau _R = \tau _A$, where the relaxation time of the Brownian noise then corresponds to the reaction time~\cite{cordes_Trouble2ndOrder_2020}. 
However, \cref{modeldef} is a genuine first order equation in the position variables $x_n$ since the second equation is the definition of the noise that does not involve the variables $x_n(t)$. 

In accordance with condition \cref{CSspe,CSacc}, it is always linearly stochastically stable, as the deterministic model is inherently stable and the noise is additive, centred, and independent of the vehicle states \cite{treiber2017intelligent}.
Such a property holds even for macroscopic traffic models \cite{Ngoduy21}, or  colored (time-correlated) noise. 
State-dependent noises, however, may impact the system linear stability, in most of the cases negatively \cite{ngoduy2019langevin,yuan2019geometric,xu2020statistical,Ngoduy21}.


\subsection{Noise-Induced Stop-and-Go Waves}

Apart from deterministic instabilities, stop-and-go dynamics can also be induced by noise.
Simulation results show that a white noise can trigger stop-and-go waves in all delayed or second-order models discussed in~\cref{sec:gen_ov}.  
The described noise-induced mechanism for stop-and-go waves is common in this class of models \cite{laval2014parsimonious,treiber2017intelligent}. However, in first-order models, white noise does not lead to stop-and-go dynamics \cite{cordes_Trouble2ndOrder_2020}.
The corresponding stop-and-go waves have less pronounced phase separation than the deterministic instabilities. 
The deterministic instabilities are more similar to vehicular traffic, while the noise-induced stop-and-go waves better describe stop-and-go waves in pedestrian dynamics, which are less pronounced.

\begin{figure}
\minipage{0.5\linewidth}
    \centering
    \includegraphics[width= 0.95 \linewidth]{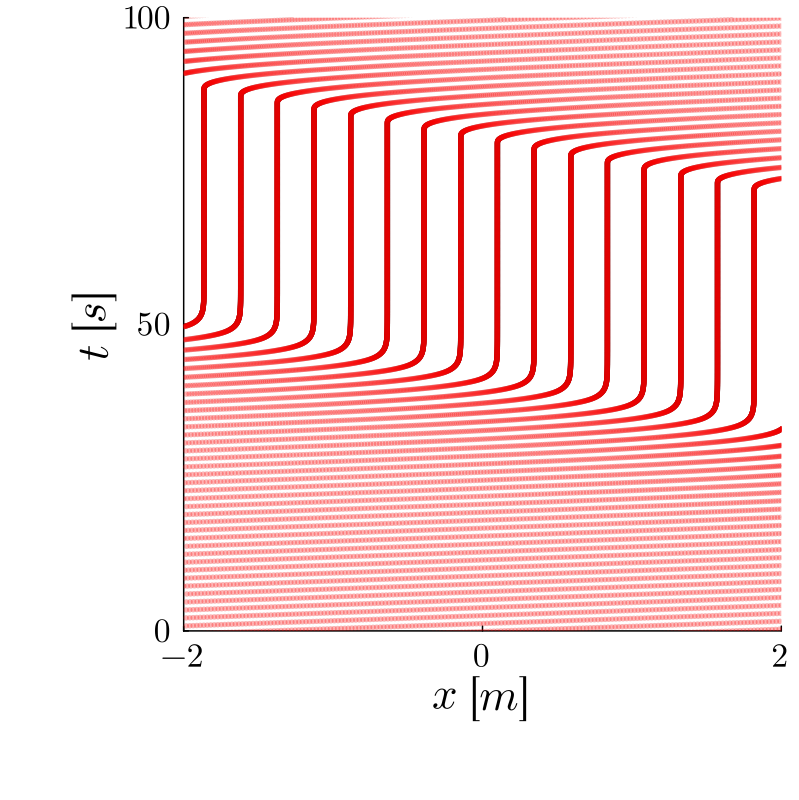}
\endminipage\hfill
\minipage{0.5\linewidth}
    \centering   
    \includegraphics[width= 0.95 \linewidth]{./Figures/noise_stopgo.pdf}
\endminipage\hfill
\caption{Illustrative scheme for the modelling of stop-and-go dynamics. Left: phase transition in the periodic solution. Right: noise-induced oscillating behaviour.}
\label{fig:noisetrans}
\end{figure}

\begin{figure}
\bigskip
\minipage{0.5\textwidth}
    \centering
    \includegraphics[width=.9 \linewidth]{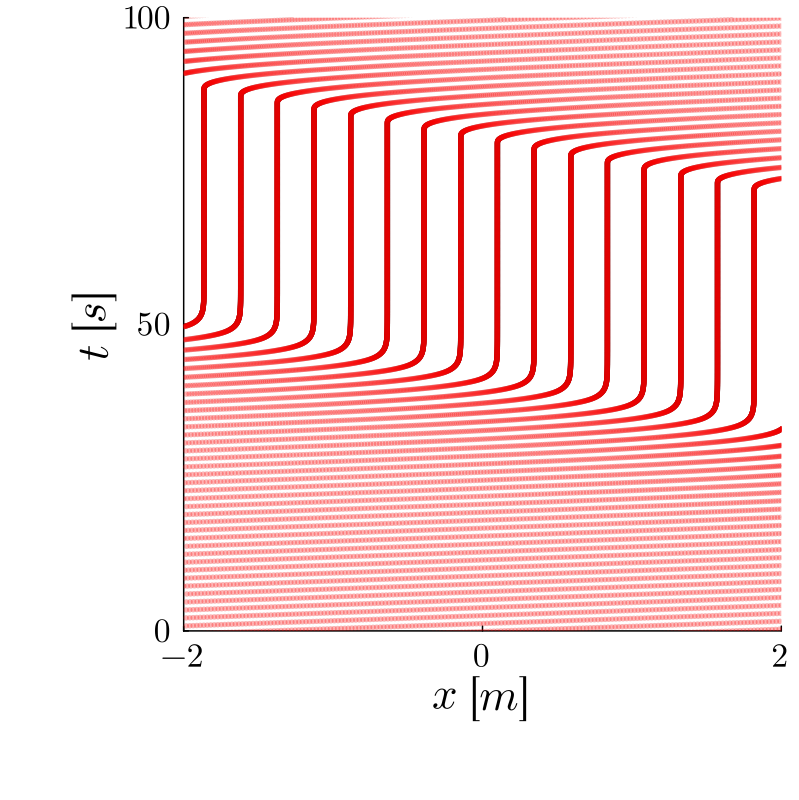}
\endminipage\hfill
\minipage{0.5\textwidth}
    \centering
    \includegraphics[width=.9 \linewidth]{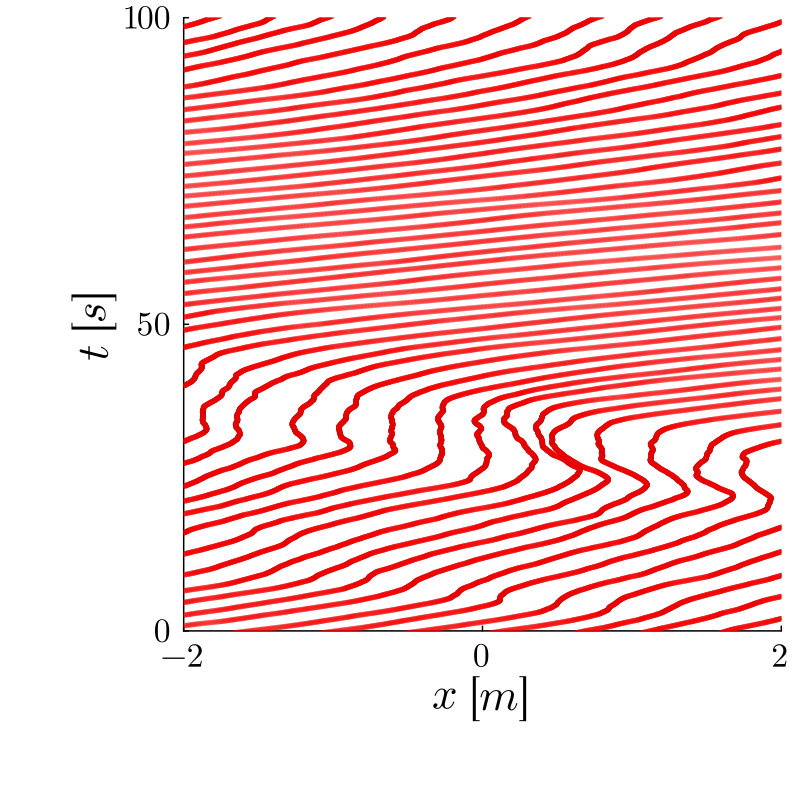}
\endminipage\hfill
\caption{Numerical trajectories of $N = 50$ agents on a ring of length $25$~m. Left: deterministic instability, i.e. the \cref{FVDM} with a non-linear optimal velocity function in the string-unstable regime. Right: noise-induced stop-and-go waves for \cref{FVDM} with a linear optimal velocity function and a white noise. Note that, this corresponds to the first-order model with a Brownian noise.}
\label{fig:traj_stop_and_go}
\end{figure}

In \cite{Tordeux2016b,tordeux_NoiseInducedStopandGoDynamics_2019}, an alternative explanation for the emergence of stop-and-go phenomenon in pedestrian flows has been proposed. This (stochastic) approach suggests that the stop-and-go waves are the result of colored noise in the dynamics of speed. 
Speed time-series exhibit a power spectrum roughly proportional to the inverse of the square noise frequency which is a typical characteristic of Brownian noise \cite{tordeux_NoiseInducedStopandGoDynamics_2019,Tordeux2016b}.
The oscillations in the system occur then in the second order, as a consequence of the perturbations introduced by the noise, particularly when the system is weakly damped. 
This mechanism differs from traditional deterministic traffic models with inertia, where stop-and-go behavior is caused by the instability of periodic dynamics through a first order phase transition. 
The fundamental distinctions between these two routes to stop-and-go behavior are summarized in~\cref{fig:noisetrans}. Representative snapshots of trajectories that display deterministic (left) and noise-induced (right) stop-and-go waves are shown in \cref{fig:traj_stop_and_go}. 
The model linearly stable as the deterministic model is inherently stable and the noise is additive, centred, and independent of the system states. 
Yet stop-and-go waves similar to those deterministic unstable models can be observed. 
The oscillating dynamics of the stochastic model results from non-linear second order effects (second order phase transition). 
Such effects can be measured by non-linear analysis of the system  variance-covariance and autocorrelation characteristics presenting oscillating features \cite{Tordeux2016b,friesen2021spontaneous}.

Two mechanisms for the description of stop-and-go waves based on relaxation processes can be identified. 
In the novel stochastic approach, the relaxation time is related to the noise and is estimated to be around 5 seconds \cite{Tordeux2016b}. This parameter corresponds to the average time period of the stochastic deviations from the phenomenological equilibrium state. 
This time can be relatively long, especially when the deviations are small and the spacings are large. 
In classical inertial approaches, the relaxation time is interpreted as reaction time and is estimated to be around 0.5 to 1 second. Technically, this parameter cannot exceed the physical time gap between the agents (around 1 to 2 seconds) without generating unrealistic behavior such as collisions, and it must be set carefully. 
Note that the class of stochastic models considered here is mainly linear. 
Noise-induced stop-and-go can also result from large perturbations in non-linear metastable systems \cite{tomer2000stable,tomer2000presence}. 
Such systems, although linearly stable, can describe deterministic stop-and-go dynamics as the noise amplitude exceeds a critical threshold.

\section{Conclusion}
\label{sec:concl}

In this contribution we have tried to give an overview and classification of following models for single-file motion. Established models from vehicular traffic modeling have been considered as well which allowed us to point out relations between the different approaches. As benchmark test we have considered single-file motion for which good empirical data is available. Despite being one-dimensional, it shows collective effects like stop-and-go motion which can provide further insight into the relevant interactions between the agents.

We have discussed the intrinsic problems of second-order models, especially those belonging to the class of social-force models, in some detail. These problems have often been reported in studies, but a systematic discussion is missing so far. As we have shown, the central issue are the effects of inertia which are much too strong, especially when the mass is identified with the mass of the agent. This leads to unrealistic behavior in certain situations which potentially could compromise simulation results, e.g. if the ordering of agents changes due to "tunneling". If this is not detected by the program, agents would no longer interact with the correct predecessor etc.

If inertia is small, first order models are a good alternative. Often they can be considered as limiting cases of second order models. We have introduced an alternative explanation for the emergence of stop-and-go waves which is not based on the classic instability mechanism. Instead the formation of these waves is triggered by correlated noise, which has also been observed empirically in time-series for the speed.

The linear stability analysis provides strong conditions for the ability of deterministic models to describe stop-and-go wave phenomena.
It remains, however, limited for stochastic models. 
Indeed, noise-induced stop-and-go dynamics operate in the second order. 
They require non-linear analysis methods, depicting oscillation behaviors in variance-covariance and auto-correlation characteristics or metastability analysis. 
Stochastic models open the door for new modelling and non-linear analysis paradigms for single-file agent dynamics that we expect to grow in the coming years.

The primary emphasis of this review centers on the single-file motion of crowds, despite the generic pedestrian motion being inherently two-dimensional. 
The rationale behind this constraint is that in such simplified scenarios, the fundamental impact of certain factors, such as inertia, can be more distinctly discerned from other influences on the motion.
As we have tried to demonstrate, this is indeed the case.
For instance the influence of inertia has been overestimated in most modeling approaches, leading to unrealistic artefacts in the depicted motion. 

However, for general applications generalizations of the models presented here are necessary. 
The aforementioned artifacts are not exclusively observed in single-file motion, but also manifest in more complex scenarios, particularly in proximity to bottlenecks.
Therefore, in a next step, it needs to be investigated how the insights gained from single-file motion can be used to improve the models such that a realistic description of pedestrian motion in general settings is obtained.
As a preliminary step, simplified two-dimensional scenarios, such as bottleneck situations or motion around corners, could be investigated. This would provide indications for the necessary modifications required in the models.


\section*{Appendix}
\setcounter{equation}{0}
\renewcommand{\theequation}{A\arabic{equation}}

\subsection*{\hypertarget{AppA}{Appendix A: Linear stability conditions for models by ordinary differential equations} \label{AppA}}

We assume in the following the existence of equilibrium states $E=(s,v)$ provided by Eqs.~(\ref{EquModSpe}) and (\ref{EquModAcc}) for first- and second-order models, respectively.
We analyse the deviation of perturbation from an equilibrium solution $E$ such that
\begin{equation}
\tilde x_i(t)=x_i(t)-x^E_i(t),\quad\text{with}\quad
\left\{\begin{array}{l}x^E_{i+1}(t)-x^E_i(t)=s,\\[1mm]
\dot x^E_i(t)=v.\end{array}\right.
\end{equation}
The equilibrium is stable if, for all agents $i$, 
\begin{equation}
    \tilde x_i(t)\rightarrow 0\quad \text{as}\quad t\rightarrow \infty
\end{equation}
in first order models, and if
\begin{equation}
    \tilde x_i(t)\rightarrow 0\quad\text{and}\quad
    \dot{\tilde x}_i(t)\rightarrow 0\quad \text{as}\quad t\rightarrow \infty
\end{equation}
in second order models.
The perturbation dynamics is then linearised using first order Taylor expansion and the model partial derivatives 
\begin{equation}
V=V(x,y),\quad a=\frac{\partial V}{\partial x}(s,v),\quad b=\frac{\partial V}{\partial y}(s,v),
\label{partialV}
\end{equation}
for speed-based models \cref{ModSpe} and
\begin{equation}
A=A(x,y,z),\quad a=\frac{\partial A}{\partial x}(s,v,v),\quad b=\frac{\partial A}{\partial y}(s,v,v),\quad c=\frac{\partial A}{\partial z}(s,v,v),
\label{partialA}
\end{equation}
for acceleration-based models \cref{ModAcc}.

\subsubsection*{First order models}

A general class of first order models introduced in \cref{ModSpe} reads
\begin{equation}
    \dot x_i(t)=V(\Delta x_i(t),\dot x_{i+1}(t)).
    \label{ModSPe2}
\end{equation}
The characteristic equation of the linearised perturbed system with $N$ agents is given by
\begin{equation}
(\lambda+a)^N-(b\lambda+a)^N=0,
\end{equation}
The local over-damped stability systematically holds for first order models while the linear string stability conditions read \cite{tordeux2012linear}
\begin{equation}
a>0,\quad b^2<1\,.
\label{CSspe}
\end{equation}


\subsubsection*{Second order models}

A general class of second order models introduced in \cref{ModAcc} is given by \begin{equation}
    \ddot x_i(t)=A(\Delta x_i(t),\dot x_i(t),\dot x_{i+1}(t))\,.
    \label{ModAcc2}
\end{equation}
The characteristic equation of the linearised perturbed system with $N$ agents reads
\begin{equation}
\big(\lambda^2-b\lambda+a\big)^N-\big(c\lambda+a\big)^N=0
\end{equation}
The linear local over-damped stability conditions reads
\begin{equation}
    a>0,\quad b<0,\quad b^2>4a
\end{equation}
while the linear string stability condition is \cite{tordeux2012linear,treiber2013traffic}
\begin{equation}
    a>0,\quad b<0,\quad b^2-c^2>2a
    \label{CSacc}
\end{equation}
Some applications for classical car-following models are provided in \cref{CSode}.
\begin{table}[!ht]
\centering
\begin{tabular}{l|c|c|c}
\multicolumn{1}{c|}{Model}&$a,b,c$&Local over-damped&String\\[1mm]
\hline&&&\\[-1mm]
OVM \cite{bando_DynamicalModelTraffic_1995}&$\displaystyle\frac1{T\tau_R},\frac{-1}{\tau_R},0$&$0<\tau_R<T/4$&$0<\tau_R<T/2$\\[-1mm]
~\cref{delayed_first_order_OV}&&\\[1mm]
FVDM \cite{jiang_FullVelocityDifference_2001}~~&~~$\displaystyle\frac1{T\tau_R},\frac{-1}{\tau_R}-\frac1{\tau_2},\frac1{\tau_2}$~~&~~$\displaystyle 0<\frac{\tau_R}{(1+\tau_R/\tau_2)^2}<T/4$~~&~~$\displaystyle 0<\frac{\tau_R\tau_2}{2\tau_R+\tau_2}<T/2$\\[-2mm]
 ~\cref{FVDM}&&\\
 with $\tau_2=\tau_RT/\tau_A$~~&&\\[1mm]
ATG \cite{tordeux2010adaptive,khound2021extending}&$\displaystyle\frac{1}{T\tau_R},\frac{-1}{T}-\frac1{\tau_R},\frac1{T}$
&$T,\tau_R>0$&$T,\tau_R>0$
\end{tabular}
\medskip
\caption{Local over-damped and string linear stability conditions for classical car-following models by second-order ordinary differential equation systems.}
\label{CSode}
\end{table}


\subsubsection*{Mixed flow models} 

For heterogeneous model parameters, i.e., 
\begin{equation}
    \ddot x_i(t)=A_i(\Delta x_i(t),\dot x_i(t),\dot x_{i+1}(t))
    \label{ModAccH}
\end{equation}
with partial derivatives $a_i$, $b_i$, $c_i$, and equilibrium state $(s_i,v)$ specific to each agent, the characteristic equation of the system with $N$ agents reads
\begin{equation}
\prod_{n=1}^N\big[\lambda^2-b_n\lambda+a_n\big]-\prod_{n=1}^N \big[c_n\lambda+a_n\big]=0.
\end{equation}
The linear string stability condition \cite{ngoduy2015effect} is given by
\begin{equation}
\forall i,~a_i>0,\quad \forall i,~b_i<0,\quad\sum_{n=1}^N\frac{b_n^2-c_n^2}{2a_n^2}\ge\sum_{n=1}^N\frac1{a_n}.
\label{CSH}
\end{equation}
This reduces to 
the classical stability condition~(\ref{CSacc}) in the homogeneous case 
($a_i=a$, $b_i=b$, and $c_i=c$ for all $i$). 
Note that the previous stability conditions \cref{CSspe,CSacc,CSH} are valid for any wavelength.


\subsubsection*{\hypertarget{AppA_kPred}{Interaction model with $K$ predecessors}  \label{AppA_kPred}}

A general class of second order interaction models with $K$ predecessors reads 
\begin{equation}
    \ddot x_i(t)=A\left(\dot x_i(t),x_{i+1}(t)-x_i(t),\dot x_{i+1}(t),\ldots, x_{i+K}(t)-x_i(t),\dot x_{i+K}(t)\right)
    \label{ModAccK}
\end{equation}
Denoting the model partial derivative as $a_k=\partial_{d_k}A$ and $b_k=\partial_{v_k}A$,
the characteristic equation in the form of a complex polynomial reads
\begin{equation}
\lambda^2=\sum_{k=1}^K a_k(e^{i2\pi lk/N}-1)+\lambda\sum_{k=0}^K b_ke^{i2\pi lk/N}.
\end{equation}
To simplify the notation, we introduce the variables
\begin{equation}
\left|~\begin{array}{l}
\mu_l=-\sum_{k=0}^K b_k\,c_{lk},\\[1mm]
\nu_l=\sum_{k=1}^K a_k\left(1-c_{lk}\right),
\end{array}\right. \qquad\quad
\left|~\begin{array}{l}
\sigma_l=-\sum_{k=1}^K b_k\,s_{lk},\\[1mm]
\rho_l=-\sum_{k=1}^K a_k\,s_{lk},
\end{array}\right.\label{pq}
\end{equation}
with $c_{lk}=\cos\left(2\pi lk/N\right)$ and $s_{lk}=\sin\left(2\pi lk/N\right)$. 
The stability condition is a generalization of the Hurwitz conditions for polynomials with complex coefficients reading \cite[Th.~3.2]{frank1946zeros}, \cite{tordeux2017influence}
\begin{equation}
\mu_l>0\quad\mbox{and}\quad
\mu_l(\nu_l\mu_l+\rho_l\sigma_l)-\rho_l^2>0,\quad\forall l.
\label{CSK}
\end{equation}
The condition is specific to each wavelength $l=1,\ldots, \lceil N/2\rceil$. 
It is general, but difficult to interpret. 
We can recover the classical stability condition \cref{CSacc} for
$K=l=1$.


\subsection*{\hypertarget{AppB}{Appendix B: Linear stability conditions for models by delay differential equations}
\label{AppB}}

For models including time delays, different stability conditions can be derived. Indeed, a delay $\tau>0$ in the dynamics generally negatively impacts the stability, see the following general stability of first- and second-order delayed models. 
We use the same notations $a$, $b$, and $c$ for the model partial derivatives as in \cref{partialV,partialA}.

\subsubsection*{First order models}
A general class of delayed first order model reads
\begin{equation}
    \dot x_i(t+\tau)=V(\Delta x_i(t),\dot x_{i+1}(t)),
    \label{ModSpeD}
\end{equation}
with $\tau\ge0$ the time delay.
The characteristic equation for the linearised system with $N$ agents is given by
\begin{equation}
(\lambda e^{\lambda\tau}+a)^N-(b\lambda+a)^N=0.    
\end{equation}
A general class of delayed first order model including an anticipation mechanism for the speed of the predecessor reads
\begin{equation}
    \dot x_i(t+\tau)=V(x_{i+1}(t)+\tau\dot x_{i+1}(t)-x_i(t+\tau),\dot x_{i+1}(t)).
    \label{ModSpeDA}
\end{equation}
Here the corresponding characteristic equation for the linearised system with $N$ agents is given by
\begin{equation}
((\lambda+a)e^{\lambda\tau})^N-(\lambda(a\tau+b)+a)^N=0.   
\end{equation}
The linear string stability conditions for these models read \cite{tordeux2012linear}
\begin{equation}
a>0,\qquad 2\tau a+b^2<1,
\label{CSspeD}
\end{equation}
for the model without anticipation \cref{ModSpeD} while the conditions is given by
\begin{equation}
a>0,\qquad (\tau a+b)^2<1,
\label{CSspeDA}
\end{equation}
for the model with anticipation \cref{ModSpeDA}. 
Note that the conditions without anticipation are more restrictive. 
Indeed, \cref{CSspeD} implies \cref{CSspeDA} but the reciprocal is false.


\subsubsection*{Second order models}

A general second order model including a delay $\tau\ge0$ reads
\begin{equation}
\ddot x_i(t+\tau)=A\big(\Delta x_i(t),\dot x_i(t),\Delta \dot x_i(t)\big),
\end{equation}
with $\Delta \dot x_i(t)=\dot x_{i+1}(t)-\dot x_i(t)$ the speed difference with partial derivative $c=\partial_z A(s,v,0)$.
The corresponding characteristic equation of the linearised system with $N$ agents is given by
\begin{equation}
(\lambda^2 e^{\lambda\tau}+\lambda(c-b)+a)^N-(\lambda c+a)^N=0   
\end{equation}
With $d=c-b$, sufficient stability conditions reads \cite{fayolle2022stability}
\begin{equation}
    a>0,\quad b<0,\quad d^2-c^2>2a,\quad\text{and}\quad \tau d<1/2.
    \label{CSacc2}
\end{equation}
The three first inequalities are identical to the stability conditions of the second order model with no delay~(\cref{CSacc2}). 
Only the last inequality relies on the delay time $\tau$. 
Note that it depends on the speed and speed difference but not on the spacing. \cref{CSdde} presents applications for selected delayed following models of the general stability conditions listed above.

\begin{table}[!ht]
\centering
\begin{tabular}{l|c|l}
\multicolumn{1}{c|}{Model}&~~Partial derivative~~&~~String stability condition\\[1mm]
\hline&\\[-1mm]
Newell model \cite{newell1961nonlinear}~(\cref{delayed_first_order_OV})~~
&~~$\displaystyle a=\frac1T$, $\beta=0$~~
&~~$\tau/T<1/2$\\[4mm]
Anticipative Newell \cite{tordeux2012linear}~~
&~~$\displaystyle a=\frac1T$, $\beta=0$~~
&~~$\tau/T<1$\\[4mm]
Delayed OV model \cite{bando1998analysis}
&~~$\displaystyle\frac1{T\tau_R},\frac{-1}{\tau_R},0$~~
&~~$\tau/T<\tau_R/(2T)<1/4$\\[4mm]
Delayed ATG model \cite{tordeux2010adaptive}~~
&~~$\displaystyle\frac{1}{T\tau_R},-\frac1{\tau_R},\frac1{T}$~~
& ~~$\tau(1/T+1/\tau_R)<1/2$\\[2.5mm]
\end{tabular}
\medskip
\caption{String linear stability conditions for classical delayed car-following models of the first order (Newell's models) and the second order (OV and ATG models).}
\label{CSdde}
\end{table}

\subsection*{\hypertarget{AppC}{Appendix C: Oscillations vs. tunneling in the social force model} \label{AppC}}
The social force model can encounter several problems, such as oscillations when the repulsive forces between pedestrians are too strong, resulting in unrealistic backwards movements and tunneling when the repulsive forces between pedestrians are too weak. 
Tunneling in single-file motion would lead to a change of order. 
The superposition principle exacerbates the aforementioned duality (oscillations vs. tunneling) by introducing a density dependency. 
To mitigate the effects of superposition, the introduction of a cut-off radius is often necessary. 
Depending on how the model is coded, this may even lead to artifacts in the overall observed dynamics.

It has been suggested in the literature that the overlapping-oscillations duality in the model can be addressed by introducing extra rules. 
One possible solution is to choose appropriate values of the repulsive forces to avoid oscillations and use an ``overlap-eliminating'' algorithm to address overlapping among pedestrians. 
Another solution proposed is to avoid overlapping by using strong repulsive forces and eliminate oscillations by setting the velocity to zero. 
However, these extra rules may add complexity to the model and are redundant as interactions among pedestrians are no longer expressed only by repulsive forces, which is in contradiction to the principle of Occam's razor. 
Additionally, it is uncertain how these modifications may affect the stability of the model, and it may make the calculations, such as those made in \cref{app_oscillator}, harder or even impossible to solve.
Therefore, it seems that the  \textit{inherent} overlapping-oscillations duality  problem of the social force model can only be solved while compromising its simplicity and properties.


\subsection*{\hypertarget{app_oscillator}{Appendix D: Damped harmonic oscillator
\label{app_oscillator}}}

The role of inertia can be understood most easily using the damped harmonic oscillator. A body or particle is subject to different forces: 1) an elastic or restoring force that is proportional to the displacement $x$ from the equilibrium position of the body and 2) a frictional damping force that is proportional to its velocity $v = \dot{x}$. The equation of motion is then given by
\begin{equation}
    m\frac{d^2x}{dt^2} + b\frac{dx}{dt} + kx = 0
    \label{eq-oscil}
\end{equation}
where $m$ is the inertial mass of the body and $b$ and $k$ are proportionality constants for the frictional and elastic force, respectively. For convenience, one introduces the relaxation time $\tau  = \frac{m}{b}$ and the free oscillation frequency $\omega_0^2 = \frac{k}{m}$.

Three different types of solutions of (\ref{eq-oscil}) exist, depending on the relative values of the proportionality constants $m$, $b$ and $k$.
\begin{enumerate}
    \item Underdamped: for $k < \frac{b^2}{4m}$ the body performs an oscillating motion with decreasing amplitude; the solution has the form $x(t) = x_0e^{-t/2\tau}\sin(\omega t+\varphi)$ where $x_0$ and $\varphi$ are determined by initial conditions. The frequency $\omega$ is smaller than the free frequency $\omega_0$ which is reached in the limit of vanishing damping ($b\to 0$).
    \item Overdamped: for $k > \frac{b^2}{4m}$ no oscillations occur and the body returns exponentially fast to the equilibrium position; the solution then has the form $x(t) = e^{-t/2\tau}\left(Ae^{\alpha t} + Be^{-\alpha t}\right)$ where $\alpha$ depends on $k,m,b$ and $A,B$ on the initial conditions.
    \item Critically damped: for $k = \frac{b^2}{4m}$ the body returns quickly to the equilibrium position ; the solution has the form $x(t) = e^{-t/2\tau}\left(A + B t\right)$ with  $A,B$ on determined by the initial conditions.
\end{enumerate}
The behavior in the overdamped regime is similar to that without inertia, i.e. for $m=0$. Therefore, the dynamics in this regime is well described by a 1st order equation.

\begin{figure}
\begin{center}
    \includegraphics[width=.65\linewidth]{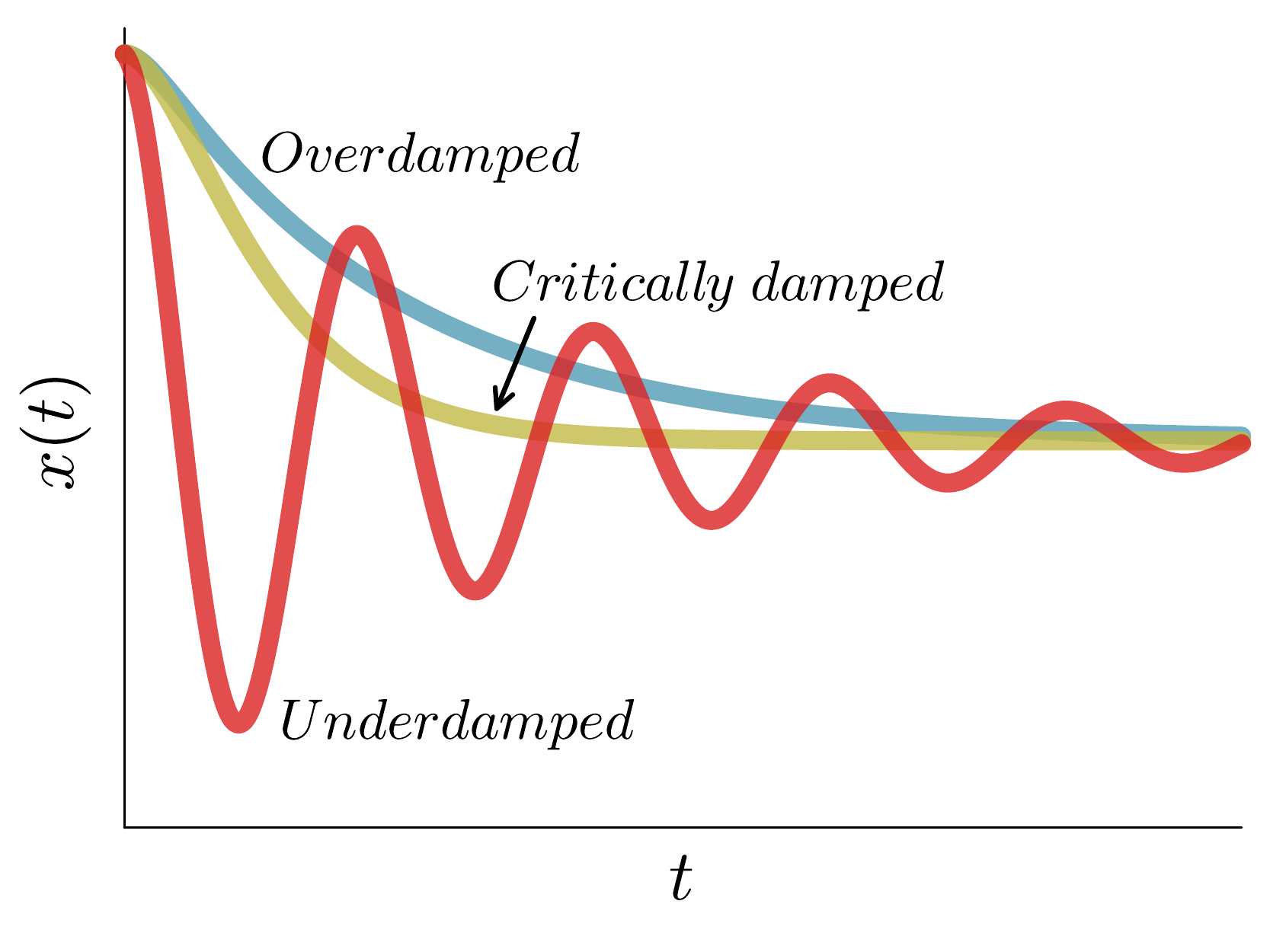}
\end{center}
\caption{Schematic representation of the behavior in the three regimes of a damped harmonic oscillator. Red: underdamped, blue: overdamped, yellow: critically damped.}
\label{fig:oscil}
\end{figure}


\paragraph{\bf Acknowledgments}  JC, MC and AT acknowledge research support from the Deutsche Forschungsgemeinschaft (DFG, German Research Foundation) – Grant number 446168800.


\end{document}